\documentclass[aps,prb,twocolumn,showpacs,superscriptaddress,groupedaddress,floatfix]{revtex4-1}
\usepackage{latexsym}
\usepackage{graphicx}
\usepackage{dcolumn}
\usepackage{bm}
\usepackage{amssymb}
\usepackage{amsmath}
\usepackage[space]{grffile}

\usepackage{subfig}

\begin{document}

\newcommand{\niceref}[1] {Eq.~(\ref{#1})}
\newcommand{\fullref}[1] {Equation~(\ref{#1})}
\title{Dynamical generation of superconducting order of different symmetries in hexagonal lattices}

\date{\today}

\begin{abstract}
The growth of superconducting order after an interaction quench in a hexagonal lattice is studied. The cases of
both time-reversal (TR) preserving graphene, as well as the TR broken
Haldane model are explored. Spin singlet superconducting order is studied where the $s$, $d+id$, and $d-id$ wave
orders are the irreducible
representations of the hexagonal lattice. For small quenches, the $d$-wave order parameter grows the fastest, a result
also expected when the system is in thermal equilibrium. For the TR symmetry preserving case, the growth rate of the two
$d$-wave orders is identical, while the TR-broken case prefers one of the chiral $d$-wave orders over the other, leading to a TR
broken topological superconductor. As the interaction quench
becomes larger, a smooth crossover is found where eventually the growth rate of the $s$-wave becomes the largest. Thus
for large interaction quenches, the $s$-wave is preferred over the $d$-wave for both TR preserving and TR broken systems.
This result is explained in terms of the high energy quasi-particles responsible for the dynamics
as the interaction quench amplitude grows. 
The results are relevant for time-resolved measurements that can probe the symmetry of the superconducting
fluctuations in a transient regime. 
\end{abstract}

\author{Hossein Dehghani}
\author{Aditi Mitra}

\affiliation{Department of Physics, New York University, 4 Washington Place, New York, NY, 10003, USA}
\maketitle
\section{Introduction}

The dynamics of how superconducting order develops, and the possibility of enhancing it under nonequilibrium conditions,
has become an active area of research spanning solid-state systems~\cite{Fausti11,Graf11,Smallwood12,Smallwood14,Beck13,Mitrano15} and
cold-atomic gases~\cite{Regal04,Zwierlein04,Bloch08,Bloch12}. These modern experiments have been complemented by intense theoretical activity
attempting to understand dynamics of superconducting systems following an interaction quench~\cite{Yin16,Yuzbashyan15,Foster15,Foster14} and
also the possibility of driving a normal system superconducting via nonequilibrium lattice vibrations~\cite{Sentef16,Knap16,Kennes17,Sentef17b}.

Most theoretical studies of nonequilibrium superconductivity, with a few exceptions~\cite{Capone15,Sentef17}, assume that only one
kind of order is relevant.
However realistic systems allow for several competing superconducting orders corresponding to
different irreducible representations of the crystal point group.
In this paper we therefore ask the following question:
is the symmetry of the superconducting order developing under nonequilibrium conditions, such as after a quench,
the same or different from the one favored by the system
in thermal equilibrium?

In this paper, we identify the dominant pairing symmetry following an interaction quench in a hexagonal lattice, both in the
presence and absence of time-reversal (TR) symmetry. For the TR-preserving system we study
doped graphene, while for the TR-broken system we consider doped graphene subjected to a high frequency circularly polarized laser. We assume
that the system before the interaction quench is in the normal state, and then study dynamics of an infinitesimal superconducting
fluctuation following the sudden switch on of
an attractive interaction. The time-evolution of the initial superconducting fluctuation is studied within linear response,
the goal being to simply identify the most unstable mode. Eventually heating effects arising due to the
energy injected in the quench, and also due to the laser, could prevent true order from developing
at the longest times. We do not address this long time behavior here. Thus our theoretical treatment
assumes collisionless or prethermal dynamics.
We are motivated by pump-probe experiments that clearly show that prethermal regimes exist, and that these transient regimes
may be unstable to ordered phases different from that in thermal equilibrium~\cite{Fausti11,Mitrano15}.

Graphene subjected to a high frequency (as compared to the bandwidth) and circularly polarized laser has been shown to be
equivalent to the Haldane model~\cite{Oka09,Kitagawa11,Haldane88}. In fact deviations from the Haldane model
are found to be small
even when accounting for how the quasi-energy bands are occupied, provided the laser is of sufficiently high frequency and weak
amplitude~\cite{Dehghani15a,Dehghani15b,Dehghani16}. Thus for this paper, we will model the TR-broken system as the Haldane model,
and study its pairing susceptiblity following an interaction quench.

Studying the onset of superconducting instabilities in hexagonal lattices is also of experimental relevance
due to the realization of such lattices using cold-atomic gases with tunable interactions~\cite{Esslinger13,Esslinger14}.
Moreover, recent experiments in bilayer graphene have shown that a strong
non-linear coupling to phonons can be achieved~\cite{Mitrano17} further increasing the possibility of inducing superconductivity.

Superconductivity in graphene has a long history~\cite{NetoRMP,Roy14}. It has been predicted
that graphene, with sufficient doping so that one is near the van-Hove singularities, can realize superconducting
order~\cite{Nandkishore12,Black14}. In particular this order can be either of the
$d_{x^2-y^2}$ or $d_{xy}$ kind.
However unlike $d_{x^2-y^2},d_{x y}$ orders that have nodes, the spontaneously TR symmetry breaking chiral $d_{x^2-y^2}\pm id_{xy}$ order
opens up a gap everywhere on the Fermi surface. Thus, from purely energetic arguments,
the preferred state in thermal equilibrium is one of the two chiral $d$ wave states~\cite{Nandkishore12,Black14}.  The $s$-wave
order, while an allowed symmetry of the lattice, is always associated with a lower critical temperature $T_c$ than the chiral $d$-wave order, so that
the latter is the preferred phase in thermal equilibrium. In the rest of the paper,
for notational convenience we will denote $d_{x^2-y^2}\pm id_{xy}$ by $d\pm id$.

The above equilibrium results for graphene
will form a backdrop for comparing our results for the superconducting order following an interaction quench.  Firstly we will show
that for the Haldane model realized from graphene subjected to a high frequency laser, the broken TR symmetry naturally
lifts the degeneracy between the two chiral modes, $d+id$ and $d-id$. Secondly, due to band flattening caused by the laser,
one need not dope the Haldane model to the same
degree as graphene to achieve the same $T_c$. Thirdly, and the main result of the paper, the interaction quench amplitude
can influence which superconducting order parameter is preferred by the system. Thus although for weak interaction quenches, the
symmetry of the order parameter is the same as in thermal equilibrium, for larger quench amplitudes, a smooth crossover to a
phase with a different superconducting order is found.

Here we discuss some subtleties both in equilibrium and out of equilibrium related to studying superconductivity in two dimensional systems.
In equilibrium, Mermin-Wagner theorem does not allow for true long range order but only quasi-long range order. Nevertheless, the
quasi-long range order is associated with correlation lengths
that are fairly long so that from the point of view of local measurements such as the spectral density, the system appears superconducting.
For this reason, the study of superconducting instabilities in two dimensional systems has been a very active area of research. In this paper, in addition
to two spatial dimensions,
we are exploring superconductivity in a transient regime where the system has not fully thermalized. 
Time-resolved measurements~\cite{Graf11,Smallwood12,Smallwood14} 
are capable of probing the symmetry of the superconducting fluctuations, and we expect the results of the paper to be relevant for such experiments. 

The paper is organized as follows. In Section~\ref{model} we describe the model, and outline the
derivation of the pairing susceptibility and the equation of motion of the
superconducting order parameter. In Section~\ref{dopedgr} we discuss the order parameter symmetries for graphene,
while in Section~\ref{dopedHaldane} we do the same for the Haldane model. We present our results for
the time evolution of superconducting fluctuation after an interaction quench in both graphene and the Haldane model in
Section~\ref{results}, and discuss the phase diagram
in terms of the interaction quench amplitude and temperature of the initial state, for a particular choice of doping.
Finally in Section~\ref{concl} we present our conclusions.

\section{Model}\label{model}

The system we will consider is a hexagonal lattice with nearest neighbor hopping, such as graphene, with and without a
circularly polarized laser. In addition we will consider interactions as our goal is to explore superconductivity.
Writing the full Hamiltonian as $H=H_0(t) + V$, where $H_0$ denotes the kinetic energy and $V$ the interactions, the kinetic part is
\begin{eqnarray}
&&H_0(t)=-t_h\sum_{k\sigma,\alpha=1,2,3}\begin{pmatrix}a_{k\sigma}^{\dagger} & b_{k\sigma}^{\dagger}\end{pmatrix}\nonumber\\
&&\times  \begin{pmatrix}0&
e^{i \vec{k}\cdot\vec{a}_{\alpha} + ia \vec{A}(t)\cdot\vec{\delta}_{\alpha}}=h_{ab}\\
e^{-i \vec{k}\cdot\vec{a}_{\alpha} - ia \vec{A}(t)\cdot\vec{\delta}_{\alpha}}=h_{ba}&0\end{pmatrix}\nonumber\\
&&\times\begin{pmatrix}a_{k\sigma} \\b_{k\sigma}\end{pmatrix}.\label{H0}
\end{eqnarray}
Above $k$ is the crystal momentum, $\sigma$ denotes the electron spin, $t_h$ is the bare tunneling amplitude between the neighboring $A,B$ sites,
$a_{k\sigma},b_{k\sigma}$ are electron annihilation operators for the $A,B$ sub-lattices respectively,
$A(t)= A_0\left[\cos(\Omega t) \hat{x} -\sin(\Omega t)\hat{y}\right]$ denotes the circularly polarized laser of amplitude $A_0$
and frequency $\Omega$.
$\vec{\delta}_i$ are the three nearest-neighbor vectors of the hexagonal lattice, which in terms of the lattice
spacing $a$ are,
\begin{eqnarray}
\!\!\vec{\delta}_1 = a\left(\frac{1}{2},\frac{\sqrt{3}}{2}\right);\vec{\delta}_2 = a\left(\frac{1}{2},-\frac{\sqrt{3}}{2}\right);
\vec{\delta}_3 = a\left(-1,0\right).
\end{eqnarray}
Above we have implicitly performed the replacement
$b_{k\sigma}\rightarrow b_{k\sigma}e^{i\vec{k}\cdot\vec{\delta}_3}$
in order to restore the periodicity of the unit-cell. Thus $\vec{a}_i$ are the translation vectors of the hexagonal lattice,
\begin{eqnarray}
&&\vec{a}_1 = \vec{\delta}_1-\vec{\delta}_3=a\left(3/2,\sqrt{3}/2\right), \nonumber\\
&&\vec{a}_2 = \vec{\delta}_2-\vec{\delta}_3=a\left(3/2,-\sqrt{3}/2\right), \nonumber\\
&&\vec{a}_3=0.\label{dela}
\end{eqnarray}

The interactions correspond to density-density interactions on sites $i$,$j$,  $Un_{i\uparrow}n_{j\downarrow}$.
A convenient way to probe the consequences of large $U$ is to map
the system to the $t$-$J$ model~\cite{Bhaskaran02,Schaffer07}
\begin{eqnarray}
&&V = J\sum_{<ij>}\biggl[\vec{S}_i\cdot\vec{S}_j -\frac{1}{4}n_in_j\biggr],\label{Veq}
\end{eqnarray}
with $J= 2t_h^2/U$.
$\vec{S}_i$ and $n_i$ are respectively the spin and number operators on site $i$, and the interactions
are only between nearest-neighbor sites. These operators act on a restricted Hilbert space that excludes doubly occupied sites.

Periodic drive and interactions will eventually lead to heating~\cite{Rigol14a}, however we are interested in very high frequency $\Omega > 6 t_h$
non-resonant driving,
where the time scales for heating processes are exponentially long~\cite{Alessio13,Abanin15,Mori16,Bukov16}. In addition, it
has been argued that an effective Hamiltonian obtained from a high-frequency expansion may be used to capture the dynamics~\cite{Kuwahara2016,Abanin17}.
With this in mind, in the
next sub-section, we derive an effective time-independent Hamiltonian by performing a high-frequency expansion. We retain
the leading term that captures the effective TR symmetry breaking due to the circularly polarized laser. This leading term opens up
a TR symmetry breaking gap at the Dirac points of graphene, mapping it to an interacting Haldane model.

\subsection{Mapping to the interacting Haldane model}
Let us Fourier transform the two off-diagonal elements of $H_0$ in Eq.~\eqref{H0},
\begin{eqnarray}
&&h_{ab}^m=\frac{1}{T}\int_0^T dt e^{-im\Omega t} h_{ab}(t) \nonumber\\
&&= i^{-m}J_{-m}(A_0a)\sum_{j=1,2,3}
e^{i \vec{k}\cdot\vec{a}_{j}+i m \alpha_j},\\
&&h_{ba}^m=\frac{1}{T}\int_0^T dt e^{-im\Omega t} h_{ba}(t) \nonumber\\
&&= (-i)^{-m}J_{-m}(A_0a)\sum_{j=1,2,3}
e^{-i \vec{k}\cdot\vec{a}_{j}+i m \alpha_j}.
\end{eqnarray}
Above $\alpha_1 = -\alpha_2 = \frac{\pi}{3}, \alpha_3 = \pi$ and $J_m$ denotes the Bessel function.
The amplitude of the laser will be given in terms of the dimensionless quantity $A_0a$, while the laser frequency
and electron temperature will be given in units of the bare tunneling amplitude $t_h$.

The high-frequency expansion for a general Hamiltonian $\hat{H}(t)$ takes the form~\cite{Eckardt15},
\begin{eqnarray}
H_{\rm eff} = \hat{H}_{m=0} + \sum_{m\neq 0} \frac{\hat{H}_m\hat{H}_{-m}}{m\hbar \Omega} + \ldots.
\end{eqnarray}
Keeping only the first two terms in the above series, the kinetic energy term becomes,
\begin{eqnarray}
&&H_{\rm eff}= -t_h
\sum_{k\sigma,\alpha=1,2,3}\begin{pmatrix}a_{k\sigma}^{\dagger} & b_{k\sigma}^{\dagger}\end{pmatrix}\nonumber\\
&&\times \begin{pmatrix}t_h\sum_{m\neq 0}
\frac{h^m_{ab}h^{-m}_{ba}}{\hbar m \Omega}& h_{ab}^{m=0}\\
h_{ba}^{m=0}& t_h\sum_{m\neq 0}\frac{h^m_{ba}h^{-m}_{ab}}{\hbar m \Omega} \end{pmatrix}\nonumber\\
&&\times \begin{pmatrix}a_{k\sigma} \\b_{k\sigma}\end{pmatrix}.
\end{eqnarray}

Using the expressions for $h_{ab}^m, h_{ba}^m$ derived above, we obtain,
\begin{widetext}
\begin{eqnarray}
&&H_{\rm eff}= -
\sum_{k\sigma}\begin{pmatrix}a_{k\sigma}^{\dagger} & b_{k\sigma}^{\dagger}\end{pmatrix}\begin{pmatrix}
2t_2\biggl[\sum_{ij=12,23,31}\sin(\vec{k}\cdot(\vec{a}_i-\vec{a}_j))\biggr]& t_1\sum_{j=1,2,3}
e^{i \vec{k}\cdot\vec{a}_{j}}\\
t_1\sum_{j=1,2,3}
e^{-i \vec{k}\cdot\vec{a}_{j}}& -2t_2\biggl[\sum_{ij=12,23,31}\sin(\vec{k}\cdot(\vec{a}_i-\vec{a}_j))\biggr]\end{pmatrix}
\begin{pmatrix}a_{k\sigma} \\b_{k\sigma}\end{pmatrix},\\
&& t_1 = t_h J_{0}(A_0a), \,\,\,\,\,\,\, \,t_2 = 2t_h^2 \sum_{m=1\ldots\infty}\frac{J_m^2(A_0a)}{\hbar m\Omega}\sin\biggl[\frac{2\pi m}{3}\biggr]\label{tren}.
\end{eqnarray}
\end{widetext}
$H_{\rm eff}$ is the Haldane model for maximal flux of $\phi=\pi/2$ threading the plaquette~\cite{Haldane88}. The above
high frequency expansion shows that in general
a periodic drive generates longer ranged matrix elements such as the next to nearest-neighbor tunneling terms
appearing on the diagonal of $H_{\rm eff}$. In addition, these tunneling matrix elements carry non-trivial phases that depend
on the polarization of the drive.
The generation of the term proportional to $\begin{pmatrix}1&0\\0&-1\end{pmatrix}$ in $H_{\rm eff}$
indicates the broken TR symmetry due to the application
of the circularly polarized laser. The high-frequency expansion is kept only up to the leading non-zero value
as it is sufficient to lift the degeneracy at the Dirac point and open up a gap that
corresponds to broken TR symmetry.
Keeping higher order terms in the frequency expansion will not generate or break any further
symmetries but will only renormalize the parameters of the Floquet Hamiltonian.

For the interaction it is sufficient to keep the $m=0$ term in the high-frequency limit.
The interaction $V$ is given by Eq.~\eqref{Veq}, which we can rewrite as,~\cite{Bhaskaran02,Schaffer07}
\begin{eqnarray}
V=-J\sum_{<ij>}h_{ij}^{\dagger}h_{ij},
\end{eqnarray}
where,
\begin{eqnarray}
h_{ij}^{\dagger}= \frac{1}{\sqrt{2}}\biggl(a_{i\uparrow}^{\dagger}b_{j\downarrow}^{\dagger}-a_{i\downarrow}^{\dagger}b_{j\uparrow}^{\dagger}\biggr),
\end{eqnarray}
is the nearest-neighbor spin-singlet creation operator. Recall that the elimination of on-site double occupancy in the $t-J$ model excludes
on site singlet formation in favor of nearest-neighbor singlet formation.  As discussed further, this nearest-neighbor spin singlet formation also gives more
structure to the singlet order-parameter, allowing for $s$,$d+id$,$d-id$ symmetries depending on how the order-parameter varies
between the three nearest-neighbor sites. 

Writing
\begin{eqnarray}
\phi_k= {\rm Arg}\biggl[\sum_{\alpha}e^{i\vec{k}\cdot\vec{a}_{\alpha}}\biggr]\label{phik},
\end{eqnarray}
it is convenient to parameterize
the non-interacting part as,
\begin{eqnarray}
&&H_{\rm eff} = \sum_{k\sigma}\begin{pmatrix}a_{k\sigma}^{\dagger} & b_{k\sigma}^{\dagger}\end{pmatrix}\nonumber\\
&&\times d_k \begin{pmatrix} \cos\theta_k & e^{i\phi_k} \sin\theta_k  \\ e^{-i\phi_k} \sin\theta_k & -\cos\theta_k\end{pmatrix}
\begin{pmatrix}a_{k\sigma} \\b_{k\sigma}\end{pmatrix}.
\end{eqnarray}
Using the pseudo-spin label $\tau=\pm$ to denote the eigenmodes of $H_{\rm eff}$ with energies $\pm |d_k|$, we use
the standard practice of parameterizing the modes by the two angles $\phi_k,\theta_k$ as follows,
\begin{eqnarray}
&&|\tau=\pm\rangle_k = \begin{pmatrix}u_{k\pm} \\v_{k\pm}\end{pmatrix},\nonumber\\
&&|\tau=+\rangle_k = \begin{pmatrix}\cos\frac{\theta_k}{2} \\ e^{-i\phi_k}\sin\frac{\theta_k}{2} \end{pmatrix} ; \,\,
|\tau=-\rangle_k = \begin{pmatrix}-\sin\frac{\theta_k}{2} \\ e^{-i\phi_k}\cos\frac{\theta_k}{2} \end{pmatrix}.\nonumber\\
\label{EVH}
\end{eqnarray}

\subsection{Quench dynamics}
We now formulate the question as follows. Let us suppose that the interactions are initially zero so that the system is in the normal phase.
An attractive interaction of magnitude $J$ is switched
on at $t=0$. As the system evolves in time under the influence of this
interaction quench, we would like to study the tendency of the system to become
superconducting. In particular, we plan to study the time-evolution of superconducting order perturbatively in the interaction.

We consider
a spin singlet superconducting order,
\begin{eqnarray}
\Delta_{\alpha} = J
\biggl\langle b_{i+\alpha\downarrow}a_{i\uparrow}-b_{i+\alpha\uparrow} a_{i\downarrow} \biggr\rangle,
\end{eqnarray}
where $\alpha=1,2,3$ denote the three nearest neighbor bonds to the site $i$. For a spatially uniform order parameter, we need not keep the
site label $i$. The explicit dependence in $\alpha$ denotes how the order-parameter can have different phases
depending on the orientation of
the nearest-neighbor bonds. This piece of information is key in order to differentiate between $s$-wave ($\Delta_{\alpha}$ independent of
$\alpha$) and a TR broken superconductor (where $\Delta_{\alpha}$ picks up different
phases for clockwise and anti-clockwise rotations around a lattice site).

After a mean-field decoupling of the interaction, we obtain,
\begin{eqnarray}
&&H(t) = H_{\rm eff}\nonumber\\
&&-\sum_{k,\alpha}\biggl[\Delta_{\alpha}(t)e^{i\vec{k}\cdot {\vec{a}}_{\alpha}}\biggl(a_{k\uparrow}^{\dagger}b^{\dagger}_{-k\downarrow}
-a_{k\downarrow}^{\dagger}b^{\dagger}_{-k\uparrow}
\biggr)+ h.c.\biggr] \label{Hmf}.
\end{eqnarray}
where $\Delta_{\alpha}(t)$ is a self-consistently determined pairing field. The mean-field decoupling neglects fluctuation terms ($\Delta_i \Delta_j$),
their role will be discussed later in the section. 

While the interaction in
Eq.~\eqref{Hmf} is written in the sub-lattice basis, it can also be written in the energy band basis
where the kinetic energy is diagonal. In that basis there are two different pairing interaction
terms, one where the electrons from the same band (intra-band) are paired, and one where
electrons from different bands (inter-band) are paired.
It is important to keep track of both the inter- and intra- band pairing amplitudes.

The equation of motion of the order parameter is,
\begin{eqnarray}
&&\Delta_{\alpha}(t)= \nonumber
J\sum_ke^{-i\vec{k}\cdot {\vec{a}}_{\alpha}}
\biggl\langle b_{-k\downarrow}(t)a_{k\uparrow}(t)-b_{-k\uparrow}(t) a_{k\downarrow}(t) \biggr\rangle,
\end{eqnarray}
where the operators obey Heisenberg time-evolution for an effectively time-dependent
Hamiltonian, $\hat{O}(t)=\tilde{T}e^{i\int_0^tdt' H(t')}\hat{O}(0) Te^{-i \int_0^tdt'H(t')}$. The time-dependence
is due to the fluctuating pairing field $\Delta_{\alpha}(t)$ in the mean-field Hamiltonian Eq.~\eqref{Hmf}.

Using Eq.~\eqref{Hmf}, and to leading order in perturbation theory in the pairing amplitude $\Delta_{\alpha}$, one need
only retain the commutator of the time-dependent pairing term with the observable. Thus one obtains, 
\begin{eqnarray}
&&\Delta_{\alpha}(t) = -iJ\sum_{kk',\beta}\int_0^t dt'\Delta_{\beta}(t')e^{-i\vec{k}\cdot {\vec{a}}_{\alpha}+ i\vec{k'}\cdot {\vec{a}}_{\beta}}\nonumber\\
&&\times \biggl\langle\biggl[\biggl(a_{k'\uparrow}^{\dagger}(t')b^{\dagger}_{-k'\downarrow}(t')
-a_{k'\downarrow}^{\dagger}(t')b^{\dagger}_{-k'\uparrow}(t')\biggr) , \nonumber\\
&&\biggl(b_{-k\downarrow}(t)a_{k\uparrow}(t)-b_{-k\uparrow}(t) a_{k\downarrow}(t) \biggr)\biggr]\biggr\rangle.
\end{eqnarray}
After performing the averages with respect to the initial state denoted by $\langle \rangle$, we obtain the equation of motion,
\begin{eqnarray}
\Delta_{\alpha}(t) = J \sum_{\beta}\int_0^t dt' \Pi^R_{\alpha\beta}(q=0,t,t') \Delta_{\beta}(t'). \label{eomDel2}
\end{eqnarray}
Above $\Pi^R$ is the response function or pairing susceptibility of the free electron gas.

Since we assume a spatially uniform order-parameter, only momentum $q=0$
component of $\Pi^R$ appears above. For a general momentum $q$, $\Pi^R$ has the form,
\begin{widetext}
\begin{eqnarray}
&&\Pi^R_{\alpha\beta}(q,t,t')=-2i\theta(t-t')\sum_{k,\tau,\tau_1=\pm}e^{-i\epsilon_{k+q\tau}(t-t')}
e^{-i\epsilon_{-k\tau_1}(t-t')}\biggl\{1-\rho_{k+q,\tau}-\rho_{-k,\tau_1}\biggr\}
\biggl[e^{-i\vec{k}\cdot ({\vec{a}}_{\alpha}-{\vec{a}}_{\beta})}|u_{k+q,\tau}|^2 |v_{-k\tau_1}|^2\nonumber\\
&&+ e^{-i\vec{k}\cdot ({\vec{a}}_{\alpha}+ {\vec{a}}_{\beta})-i \vec{q}\cdot {\vec{a}}_{\beta}}u_{k+q\tau}v_{k+q\tau}^*
v_{-k\tau_1}u^*_{-k\tau_1}\biggr],\nonumber\\\label{PiRHaldane}
\end{eqnarray}
\end{widetext}
where $\rho_{k,\tau}$ is the occupation probability of the $k,\tau$ level in the absence of interactions.

We now briefly discuss the relation of Eq.~\eqref{eomDel2} to
what is usually done in equilibrium to determine the superconducting phase. In equilibrium, $t=\infty$, and going into frequency space, one solves
$\Delta_{\alpha} = J \sum_{\beta}\Pi^R_{\alpha \beta}(q=0,\omega=0)\Delta_{\beta}$. Secondly, the response function $\Pi^R$ is evaluated for
the full interacting electron gas in the normal phase. Thus the only assumption is that the pairing amplitude $\Delta_{\alpha}$
is small. Beyond this, an additional assumption can be made, and that is of weak $J/t_h$. This is the BCS approximation, where now
$\Pi^R$ is evaluated for the free Fermi gas.

In our problem, namely a quench, since the average is always with respect to the initial state which is that of a free Fermi gas, the
linear-response assumption is tied to perturbation theory in $J$ as well. Strictly speaking this assumption will break down when quartic
terms  ($\Delta_{i}\Delta_j$) or inelastic scattering between particles become important. 
The time scale for this from Fermi's Golden rule, in units of the hopping
$t_h$, is $ t_{\rm in} \sim 1/(J^2T)$, where $T$ is the temperature of the electron gas, and $J,T$ are in units of the hopping $t_h$. 
At times longer than this time, thermalizing processes will become active. 

Since finite temperatures are not detrimental to superconductivity in spatial dimensions $d>1$, including $d=2$ provided $T<T_{\rm BKT}$,
where $T_{\rm BKT}$ is the Brezenskii Kosterlitz Thouless temperature, our conclusions are not completely invalidated in the
inelastic scattering dominated regime. We discuss this issue in more detail in Section~\ref{results}.

Mean-field~\cite{Yuzbashyan15,Capone15} and linear instability analyses~\cite{Knap16},  
despite their apparent simplicity, are relevant to 
experiments that can probe the short time regime where 
quasi-particles have formed even though the bulk system has not developed a traditional Meissner effect. In particular time-resolved ARPES
can pick out the dispersion of the quasi-particles, and hence the symmetries of the order-parameter~\cite{Graf11,Smallwood12,Smallwood14}. 

In what follows we will assume that initially the free fermions were in equilibrium at temperature $T$
and chemical potential $\mu$. Thus for a free fermion dispersion $\epsilon_k$ which is also inversion symmetric,
$\epsilon_{k,\tau} = \epsilon_{-k,\tau}$, we obtain,
\begin{eqnarray}
&&1-\rho_{k,\tau}-\rho_{-k,\tau_1} = \delta_{\tau,\tau_1}\tanh\left(\frac{\epsilon_{k,\tau}+\mu}{2T}\right)\nonumber\\
&&+\delta_{\tau,-\tau_1}\frac{1}{2}\biggl[\tanh\left(\frac{\epsilon_{k,\tau}+\mu}{2T}\right)-
\tanh\left(\frac{\epsilon_{k,\tau}-\mu}{2T}\right)\biggr].
\end{eqnarray}
A given temperature $T$ and chemical potential $\mu$ denotes a doping $\delta$ away from
half-filling,
\begin{eqnarray}
\delta = \frac{1}{2}\sum_k\biggl[\tanh\biggl(\frac{\epsilon_k+\mu}{2T}\biggr)
-\tanh\biggl(\frac{\epsilon_k-\mu}{2T}\biggr)\biggr].
\end{eqnarray}
We will present results for a non-zero doping $\delta=0.1$. For zero doping, since the chemical potential is in the gap in equilibrium, 
the density of states at the Fermi energy vanishes and superconductivity does not occur in the BCS limit.

In equilibrium, in order to determine the critical temperature for some coupling $J$, one solves the gap equation, $J\Pi^R(q=0,\omega=0)=1$, where,
\begin{eqnarray}
&&\Pi^R_{\alpha \beta}(q,\omega=0)= \sum_{k,\tau,\tau_1=\pm}\frac{1}{\epsilon_{k+q\tau}+ \epsilon_{-k\tau_1}}\nonumber\\
&&\times \biggl\{1-\rho_{k+q,\tau}-\rho_{-k,\tau_1}\biggr\}\biggl[e^{-i\vec{k}\cdot ({\vec{a}}_{\alpha}-{\vec{a}}_{\beta})}|u_{k+q,\tau}|^2 |v_{-k\tau_1}|^2
\nonumber\\
&&+ e^{-i\vec{k}\cdot ({\vec{a}}_{\alpha}+ {\vec{a}}_{\beta})-i \vec{q}\cdot {\vec{a}}_{\beta}}u_{k+q\tau}v_{k+q\tau}^*
v_{-k\tau_1}u^*_{-k\tau_1}\biggr]. \label{PiRHG}
\end{eqnarray}
In the next sections, we will directly solve Eq.~\eqref{eomDel2} for some small, initial randomly chosen $\Delta_{\alpha}$. Since
$\alpha$ takes three values, we denote the superconducting order as a vector $\vec{\Delta}$.
Growing and decaying solutions in time will indicate whether the system is susceptible to pairing. In addition, how the three
components $\Delta_{\alpha=1,2,3}$ of the order parameter grow will indicate
the preferred symmetry of the superconducting order parameter.

It might seem that the only place the chemical potential appears is in the initial distribution function. However
we will measure all energies with respect to this chemical potential.
This choice is equivalent to multiplying the superconducting order-parameter by the phase
$e^{-2i\mu t}$.
Thus in Eq.~\eqref{PiRHaldane} we will
denote the energy of the upper band $\epsilon_{k+}=\epsilon_k+\mu$,
and that of the lower band as $\epsilon_{k-}=-\epsilon_k+\mu$, where we have also used the fact that particle-hole symmetry in the problem
makes the band dispersions symmetric about zero energy.

Although Eq.~\eqref{eomDel2} was derived perturbatively in $J$, yet the solution of the integral equation is
non-perturbative in $J$. To see this note that if initially at $t=0$ we had a small seed order parameter $\delta(t')\Delta_0$, then the first
order correction from Eq.~\eqref{eomDel2} is $\Delta_1(t) = J\Pi^R(t,0)\Delta_0$. The second order correction is obtained from substituting this
back in Eq.~\eqref{eomDel2} to obtain $\Delta_2(t) = J^2\int dt_1\Pi^R(t,t_1)\Pi^R(t_1,0)\Delta_0$, and so on. Thus the solution for Eq.~\eqref{eomDel2}
can be recast as an integral equation $\Delta(t)= D^{-1}\Delta_0$, essentially the Dyson equation in real time, where $(1-J\Pi)=D$. This leads to
a solution, which after a short transient $t <O(1)$ stabilizes to an exponentially growing or decaying solution, $\Delta \sim e^{(J-J_c)t}$ where $J_c$ is 
a critical coupling. The appearance of $J$ in the argument of the exponential shows the non-perturbative role of $J$ in determining the superconducting phase.

From complex analysis, the relation $\Delta(t)= D^{-1}\Delta_0$ implies that locations of
zeros of $D$ in the complex frequency plane will determine the growth or decay rate of the order parameter.
Thus for later discussions we will find it convenient to interpret the results of the time evolution in
terms of the location of the zeros of $D$ in the complex frequency (denoted by $z$) plane. The explicit form for $D(q,z)$ is,
\begin{eqnarray}
&&D_{\alpha\beta}(q,z) = \delta_{\alpha\beta}- J\Pi^R_{\alpha \beta}(q,z)\nonumber\\
&&= \delta_{\alpha\beta}- 2J\sum_{k,\tau,\tau_1=\pm}\biggl\{\frac{1-\rho_{k+q,\tau}-\rho_{-k,\tau_1}}{\epsilon_{k+q\tau}+ \epsilon_{-k\tau_1}-2iz}\biggr\}\nonumber\\
&&\times \biggl[e^{-i\vec{k}\cdot ({\vec{a}}_{\alpha}-{\vec{a}}_{\beta})}|u_{k+q,\tau}|^2 |v_{-k\tau_1}|^2\nonumber\\
&&+ e^{-i\vec{k}\cdot ({\vec{a}}_{\alpha}+ {\vec{a}}_{\beta})-i \vec{q}\cdot {\vec{a}}_{\beta}}u_{k+q\tau}v_{k+q\tau}^*
v_{-k\tau_1}u^*_{-k\tau_1}\biggr]. \label{Dz}
\end{eqnarray}

In section~\ref{dopedgr} we discuss the symmetries of the superconducting order for graphene while in section~\ref{dopedHaldane}
we discuss the same for the Haldane model. The results of the time-evolution are presented in Section~\ref{results}.

\section{Pairing symmetries of graphene} \label{dopedgr}

Since graphene has only nearest neighbor hopping, $t_2=0,t_1=t_h$, the modes are,
\begin{eqnarray}
&&|\tau=\pm\rangle_k = \begin{pmatrix}u_{k\pm} \\v_{k\pm}\end{pmatrix},\nonumber\\
&&|\tau=+\rangle_k = \frac{1}{\sqrt{2}}\begin{pmatrix}1 \\ e^{-i\phi_k}\end{pmatrix} ;
|\tau=-\rangle_k = \frac{1}{\sqrt{2}}\begin{pmatrix}-1 \\ e^{-i\phi_k} \end{pmatrix}.\nonumber
\end{eqnarray}
$\phi_k$ is defined in Eq.~\eqref{phik}.

Since under $k\rightarrow -k$, $H_k \rightarrow \tau_xH_k\tau_x$, then,
\begin{eqnarray}
u_{k\tau} = u_{-k\tau}^*, v_{k\tau}= v_{-k\tau}^*.
\end{eqnarray}
This gives the following expression for the pair susceptibility,
\begin{widetext}
\begin{eqnarray}
&&J\Pi^R_{\alpha \beta}(t,t')=2i\theta(t-t')\frac{J}{4}\sum_k\biggl[e^{-2i(\epsilon_k+\mu)(t-t')}
\tanh\biggl(\frac{\epsilon_k+\mu}{2T}\biggr)
+e^{-2i(-\epsilon_k+\mu)(t-t')} \tanh\biggl(\frac{-\epsilon_k+\mu}{2T}\biggr)\biggr]\nonumber\\
&&\times \biggl[\cos\biggl(\vec{k}\cdot ({\vec{a}}_{\alpha}-{\vec{a}}_{\beta})\biggr)
+ \cos\biggl(\vec{k}\cdot ({\vec{a}}_{\alpha}+ {\vec{a}}_{\beta})-2\phi_k\biggr)\biggr] \nonumber\\
&&+ i\theta(t-t')e^{-2i\mu(t-t')}\frac{J}{2}\sum_k\biggl[\frac{\sinh(\frac{\mu}{T})}{\cosh{(\frac{\epsilon+\mu}{2T})}\cosh{(\frac{\epsilon-\mu}{2T})}}\biggr]
\biggl[\cos\biggl(\vec{k}\cdot ({\vec{a}}_{\alpha}-{\vec{a}}_{\beta})\biggr)
- \cos\biggl(\vec{k}\cdot ({\vec{a}}_{\alpha}+ {\vec{a}}_{\beta})-2\phi_k\biggr)\biggr].
\end{eqnarray}
\end{widetext}
The dc component of the response function $\Pi^R(\omega=0)$ agrees with Ref.~\onlinecite{Black14}, where the first
line above corresponds to intra-band pairing, and the second line to inter-band pairing.

Let us now discuss the symmetries of the $\Pi^R$ matrix.
For every $\vec{k}$, there are two others oriented by $2\pi/3, 4\pi/3$ from
it. We label the triad as,
\begin{eqnarray}
&&\vec{k}_{m=1,2,3} \nonumber\\
&&= k\left[\cos\biggl(\theta_k+(m-1)\frac{2\pi}{3}\biggr), \sin\biggl(\theta_k+(m-1)\frac{2\pi}{3}\biggr)\right].\nonumber
\end{eqnarray}
We label the nearest-neighbor vectors $\vec{\delta}_{\alpha}$ similarly,
\begin{eqnarray}
&&\vec{\delta}_{j=1,2,3} \nonumber\\
&&= a \left[\cos\biggl(\frac{\pi}{3}-(j-1)\frac{2\pi}{3})\biggr), \sin\biggl(\frac{\pi}{3}-(j-1)\frac{2\pi}{3}\biggr)\right].\nonumber
\end{eqnarray}
Thus,
\begin{eqnarray}
\vec{k}_m\cdot\vec{\delta}_j = ka\cos\biggl(\theta_k + (m+j)\frac{2\pi}{3}+\frac{\pi}{3}\biggr).
\end{eqnarray}
From above we see $\vec{k}_m\cdot\vec{\delta}_j= \vec{k}_{m\pm1}\cdot\vec{\delta}_{j\mp1}$. In addition if $j$ or $m$ change by $\pm 3$ or its multiples,
the function comes back to itself.
This implies the following relations,
\begin{eqnarray}
&&\phi(k_m) = {\rm Arg}\biggl[\sum_{i=1,2,3} e^{i \vec{k}_m\cdot\vec{\delta}_i}\biggr] \Rightarrow \phi(k_1) \!= \!\phi(k_2)=\! \phi(k_3),\nonumber\\
&&m(k_m) = {\rm Abs}\biggl[\sum_{i=1,2,3} e^{i \vec{k}_m\cdot\vec{\delta}_i}\biggr] \Rightarrow m(k_1) \!= \!m(k_2)=\! m(k_3),\nonumber\\
&&\sum_{j=1,2,3}m(k_m)f(\vec{k}_m\cdot\vec{\delta}_j-\phi_{k_m}) = f'_m\Rightarrow f'_1=f'_2=f'_3,\nonumber\\
&&\sum_{\vec{k}}m(\vec{k})f(\vec{k}\cdot\vec{\delta}_j-\phi_k)=g_j \Rightarrow g_1=g_2=g_3.
\end{eqnarray}
The above equalities simply reflect the $C_3$ symmetry of the hexagonal lattice. While the first two equalities
above directly influence the energy eigenvalues and eigenvectors, the diagonal component of the pairing susceptibility
$\Pi_{\alpha\alpha}$ is of the last form. Note that the vectors $\vec{\delta}, \vec{a}$ are related by a constant shift (see Eq.~\eqref{dela}), 
thus the arguments for the symmetries of the $\Pi$ matrix hold irrespective of whether the $\Pi$ matrix
is written in the $\vec{\delta}$ or $\vec{a}$ basis.
 
Thus the $C_3$ symmetry implies $\Pi_{11}=\Pi_{22}=\Pi_{33}=A$.
Other components of $\Pi$ may be written as,
\begin{eqnarray}
\sum_{\vec{k}}m(\vec{k})l(\vec{k}\cdot\vec{\delta}_j-\phi_k)l(\vec{k}\cdot\vec{\delta}_i-\phi_k)=L_{ji}= L_{ij}.
\end{eqnarray}
The above implies $\Pi_{\alpha\beta}=\Pi_{\beta\alpha}$.  This together with the fact that $L_{ji}= L_{j\pm 1,i\pm 1}$ gives
$\Pi_{12}= \Pi_{23}=\Pi_{31}=B$ and $\Pi_{12}=\Pi_{13}=\Pi_{23}=B$.

In fact the above symmetries hold at any instant of time $t$, so that $\Pi^R$ has the general form,
\begin{eqnarray}
\Pi^R(t) = \begin{pmatrix}A(t) &B(t) & B(t) \\B(t) &A(t) & B(t) \\B(t) &B(t) &A(t)\label{PiGs}
\end{pmatrix}.
\end{eqnarray}

From the structure of Eq.~\eqref{PiGs} it follows that the eigenvalues and eigenvectors of the $\Pi^R$ matrix at any instant of time are
\begin{eqnarray}
&&\lambda_s(t) = \Pi_{11}(t) + 2\Pi_{12}(t);\,\,\,\,\,\, \Delta_s = \frac{1}{\sqrt{3}}\left[1,1,1\right],\\
&&\lambda_{d_1}(t) = \lambda_{d_2}(t)= \Pi_{11}(t) - \Pi_{12}(t), \nonumber\\
&&\Delta_{d_1} = \frac{1}{\sqrt{6}}\left[2,-1,-1\right]; \,\,\,\,\,\, \Delta_{d_2} = \frac{1}{\sqrt{2}}
\left[0,1,-1\right].
\end{eqnarray}
While the eigenvalues depend explicitly on time, the eigenvectors do not. Thus all throughout the time-evolution,
there are three mutually orthogonal directions that remain the same.
These directions correspond to a non-degenerate $s$-wave solution and a doubly degenerate $d$-wave solution. Any random initial condition
for the superconducting order parameter evolves independently along these three directions. We track the time-evolution,
and the fastest growing order parameter determines the nature of the superconductor at steady state. Due to the degeneracy of the two
$d$-wave modes, the growth rate of the two $d$-wave orders, or any linear combination of the two $d$-wave orders will be identical.

Here we should mention that in equilibrium, the $T_c$ of the superconducting phase and its symmetry is determined from $1-J\Pi^R(\omega=0)=0$.
This treatment~\cite{Black14} gives the largest eigenvalue, and therefore the dominant instability to correspond to
$d$-wave. Since this is doubly degenerate, the order parameter symmetry is not uniquely determined. Instead,
energetic considerations indicate that the TR breaking combination $d_{x^2-y^2}+id_{xy}=\frac{1}{\sqrt{3}}\biggl[1,e^{2\pi i/3},e^{4\pi i/3}\biggr]$
(or its complex conjugate $d_{x^2-y^2}-id_{xy}$) will be favored. This is because the
chiral order parameter has no nodes.

For the case of the quench, such energy minimization
considerations no longer hold. Instead if the time-evolution shows the $d$-wave to be the fastest growing order parameter,
the precise order parameter could be any linear combination in the degenerate sub-space. If the fastest growing order parameter is the $s$-wave, then
the order parameter symmetry is uniquely determined. For the Haldane model, as we discuss below, the breaking of TR lifts the degeneracy in the
$d$-wave sub-space.

\section{Pairing symmetries of the Haldane model} \label{dopedHaldane}
We now discuss the pairing susceptibility of the Haldane model.  We again make an assumption that the fermions
are in thermal equilibrium at temperature $T$ and chemical potential $\mu$ before the interaction quench.
This gives,
\begin{widetext}
\begin{eqnarray}
&&\Pi^R_{\alpha\beta}(q=0,t,t')=\nonumber\\
&&-2i\theta(t-t')\sum_{k}e^{-2i(\epsilon_{k}+\mu)(t-t')}
\biggl\{\tanh\left(\frac{\epsilon_k+\mu}{2T}\right)\biggr\}
\biggl[e^{-i\vec{k}\cdot ({\vec{a}}_{\alpha}-{\vec{a}}_{\beta})}\cos^4\frac{\theta_k}{2}
+ e^{-i\vec{k}\cdot ({\vec{a}}_{\alpha}+ {\vec{a}}_{\beta})}e^{2i\phi_k}\cos^2\frac{\theta_k}{2}\sin^2\frac{\theta_k}{2}\biggr]\nonumber\\
&&-2i\theta(t-t')\sum_{k}e^{-2i(-\epsilon_{k}+\mu)(t-t')}
\biggl\{\tanh\left(\frac{-\epsilon_k+\mu}{2T}\right)\biggr\}
\biggl[e^{-i\vec{k}\cdot ({\vec{a}}_{\alpha}-{\vec{a}}_{\beta})}\sin^4\frac{\theta_k}{2}
+ e^{-i\vec{k}\cdot ({\vec{a}}_{\alpha}+ {\vec{a}}_{\beta})}e^{2i\phi_k}\cos^2\frac{\theta_k}{2}\sin^2\frac{\theta_k}{2}\biggr]\nonumber\\
&&-2i\theta(t-t')\sum_{k}e^{-2i\mu(t-t')}
\biggl\{\tanh\left(\frac{\epsilon_k+\mu}{2T}\right)+ \tanh\left(\frac{-\epsilon_k+\mu}{2T}\right)\biggr\}
\biggl[e^{-i\vec{k}\cdot ({\vec{a}}_{\alpha}-{\vec{a}}_{\beta})}\cos^2\frac{\theta_k}{2}\sin^2\frac{\theta_k}{2}\nonumber\\
&&\,\,\,\,\,\,\,\,\,\,\,\,\,\,\,\,\,\,\,\,\,\,\,\,\,\,\,\,\,\,\,\,\,\,\,\,\,\,\,\,\,\,\,\,\,\,\,\,\,\,\,\,\,\,\,\,\,\,\,\,\,\,\,\,\,\,\,\,\,\,\,\,\,\,\,
\,\,\,\,\,\,\,\,\,\,\,\,\,\,\,\,\,\,\,\,\,\,\,\,\,\,\,\,\,\,\,\,\,\,\,\,\,\,\,\,\,\,\,\,\,\,\,\,\,\,\,\,\,\,\,\,\,\,\,\,\,\,\,\,\,\,\,\,\,\,\,\,\,\,\,
\,\,\,\,\,\,\,\,\,\,\,\,\,\,\,\,\,\,\,\,\,\,\,\,\,
- e^{-i\vec{k}\cdot ({\vec{a}}_{\alpha}+ {\vec{a}}_{\beta})}e^{2i\phi_k}\cos^2\frac{\theta_k}{2}\sin^2\frac{\theta_k}{2}\biggr].
\end{eqnarray}
\end{widetext}

Like graphene, the Haldane model also has $C_3$ symmetry associated with
invariance under rotations by $2\pi/3$. Thus following the arguments given for
graphene in the previous section, we conclude,
$\Pi_{11}=\Pi_{22}=\Pi_{33}=A$, $\Pi_{12}=\Pi_{23}=\Pi_{31}=B$, and
$\Pi_{13}=\Pi_{21}=\Pi_{32}=C$. Thus we may write,
\begin{eqnarray}
\Pi^R(t) = \begin{pmatrix}A(t) &B(t) & C(t) \\C(t) &A(t) & B(t) \\B(t) &C(t) &A(t)
\end{pmatrix}.
\end{eqnarray}
Unlike graphene however, $\Pi_{\alpha\beta}\neq \Pi_{\beta \alpha}$, when $ \alpha\neq \beta$.
In the dc limit $\Pi^R(\omega=0)$ is Hermitian forcing $C(\omega=0)=B^*(\omega=0)$.

The eigenvalues and eigenvectors of the $\Pi(t)$ matrix are
\begin{eqnarray}
&&\lambda_s(t) = A(t)+B(t)+C(t);\,\, \Delta_s = \frac{1}{\sqrt{3}}\left[1,1,1\right],\\
&&\lambda_{d+id}(t) = A(t) + B(t) e^{2\pi i/3} + C(t) e^{4\pi i/3} ,\nonumber\\
&&\Delta_{d+id} = \frac{1}{\sqrt{3}}\begin{pmatrix}1\\ e^{2\pi i/3}\\e^{4\pi i/3}\end{pmatrix},\\
&&\lambda_{d-id}(t)=A(t) + B(t) e^{4\pi i/3} + C(t) e^{2\pi i/3},\nonumber\\
&&\Delta_{d-id} =\frac{1}{\sqrt{3}}\begin{pmatrix}1\\ e^{4\pi i/3}\\e^{2\pi i/3}\end{pmatrix}.
\end{eqnarray}
Graphene corresponds to the case where $\Pi_{\alpha\beta}=\Pi_{\beta\alpha}$ making $B=C$ and $\lambda_{d+id}=\lambda_{d-id}=A-B$.
Thus we find that there are no degenerate eigenvalues for the Haldane model as broken TR prefers one of the chiral $d$-wave modes over the other.

Just as for graphene, here too the eigenvalues of $\Pi^R$ are time-dependent, while the mutually orthogonal directions stay static. We
again study the time-evolution of an initial random but small vector $\vec{\Delta}$, project the evolution along the three mutually
orthogonal directions corresponding to the
eigenvectors of $\Pi^R$, and determine the phase from the fastest growing mode.

Before we present the results for the quench, we point out that in equilibrium, the Haldane model obtained from applying a high frequency
circularly polarized laser to graphene,
is more susceptible to pairing than graphene. This is because the laser flattens out the band
as can be seen from the renormalization of the hopping amplitudes from $t_h \rightarrow t_h J_0(A_0a)$ in Eq.~\eqref{tren}.
As a consequence, the same $T_c$ can be
obtained at much lower doping levels in the presence of the laser than in graphene.
As an example, the pairing susceptibilities derived above give that graphene for $J=0.6 t_h$ has a $T_c=0.01t_h$ at a doping of $\delta=0.11$.
In contrast the Haldane model
realized from a laser of amplitude $A_0a=1.0$ and frequency $\Omega =10 t_h$ shows the same $T_c$ but at a much lower doping of $\delta \sim 0.01$.
\begin{figure}
\includegraphics[width = 1.0\columnwidth]{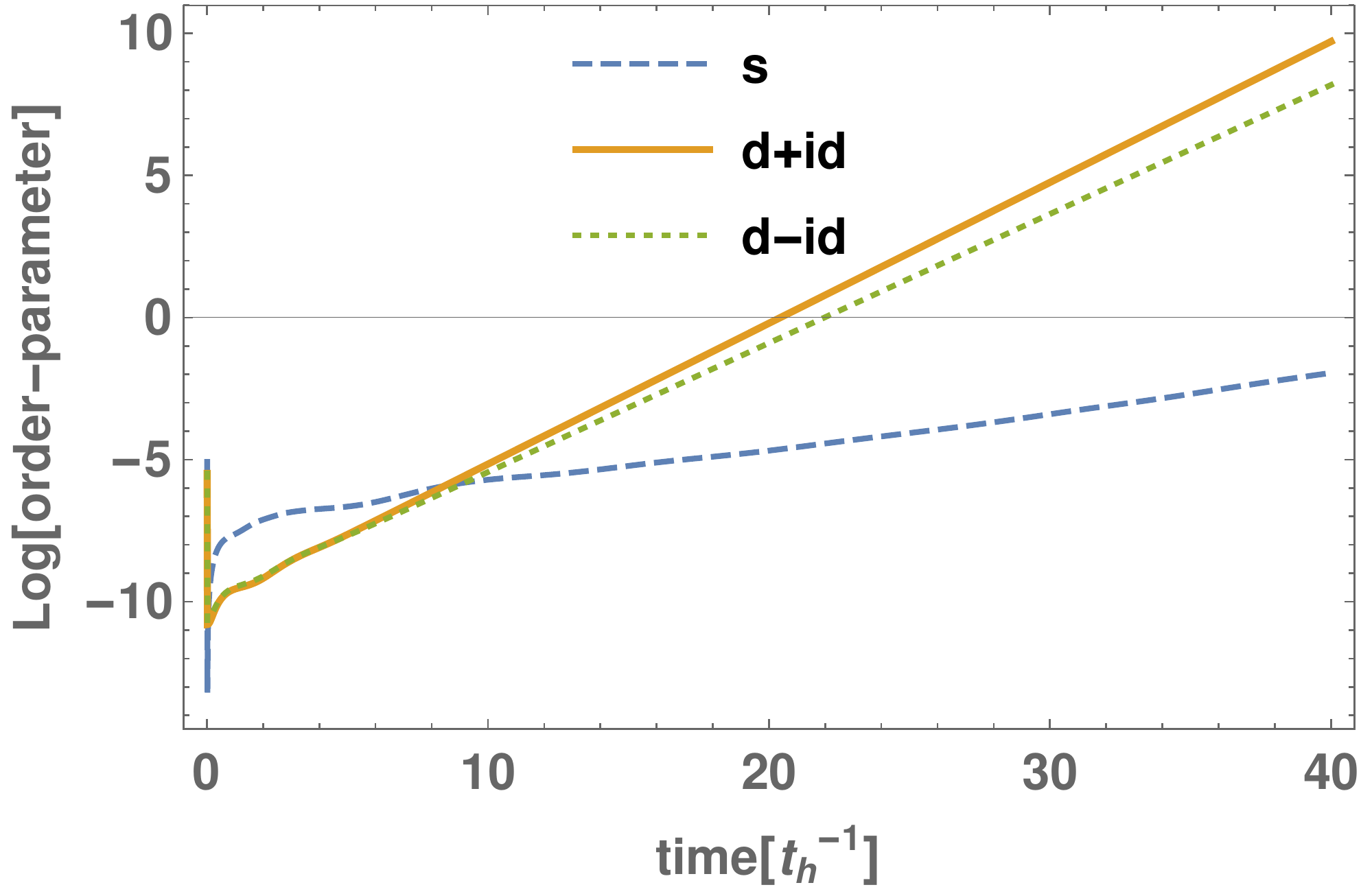}
\caption{
Haldane model $A_0a=0.5,\Omega=10 t_h, J=1.82t_h, T=0.01 t_h$ and doping $\delta=0.1$. Time-evolution of the logarithm of an
initial random vector. The time-evolution is projected along the three orthogonal directions
with $s$, $d+id$, $d-id$ symmetry. The slopes indicate that for the chosen parameters $d+id$ is the fastest growing instability,
followed by $d-id$ and then $s$. Time is in units of $t_h^{-1}$.
}
\label{fig1}
\end{figure}

\begin{figure}
\includegraphics[width = 1.0\columnwidth]{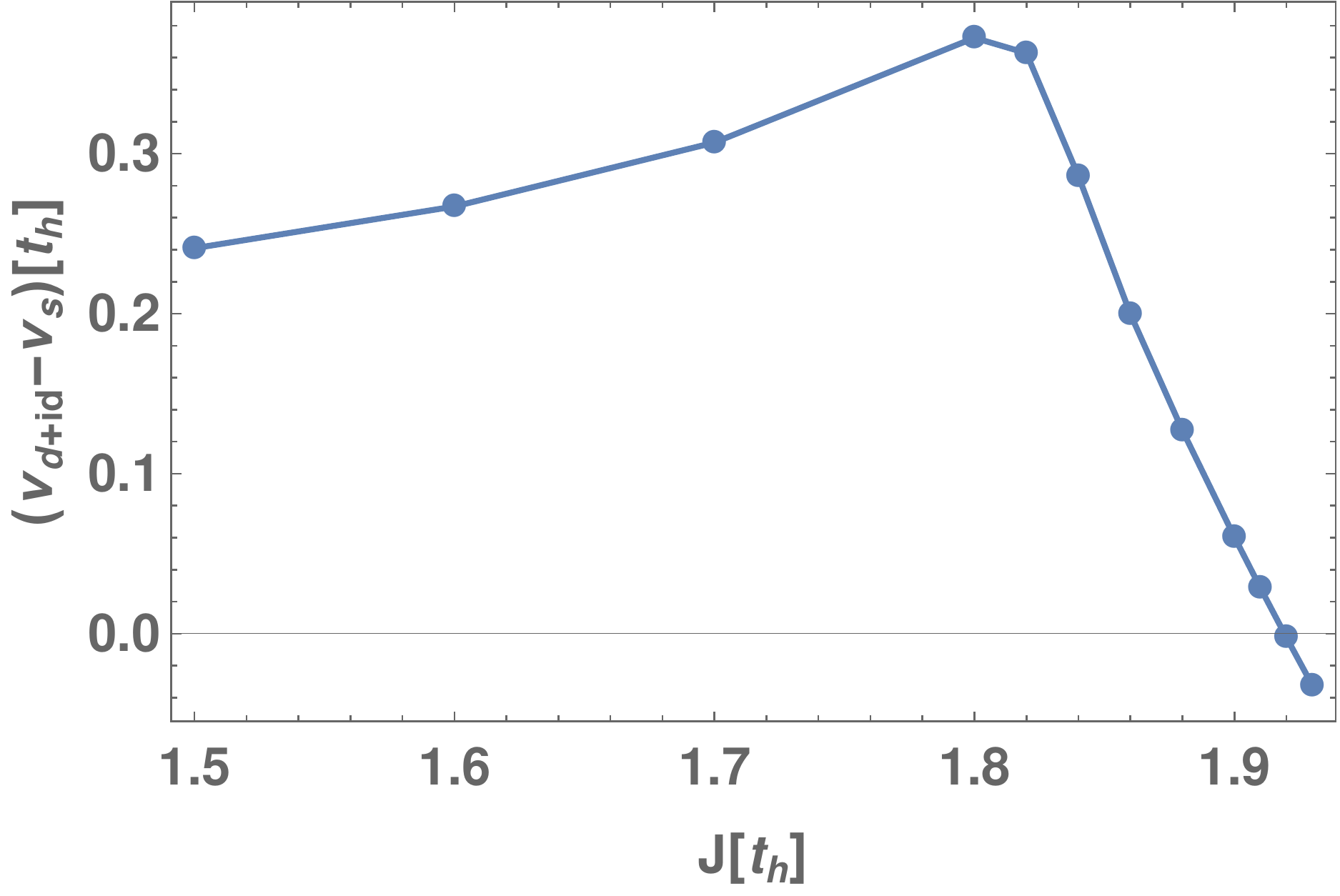}
\caption{
Haldane model $A_0a=0.5,\Omega=10t_h,T=0.01t_h$ and doping $\delta=0.1$  (same as Fig.~\ref{fig1}). As the quench amplitude $J$ is increased,
the difference between the growth rate of chiral $d$-wave ($\nu_{d+id}$) and $s$-wave ($\nu_s$) varies as shown above. The difference first
increases, and then decreases rapidly. For quench amplitudes larger than $J_c\sim 1.9 $ the $s$-wave is preferred.
}
\label{fig2}
\end{figure}

\begin{figure}
\includegraphics[width = 1.0\columnwidth]{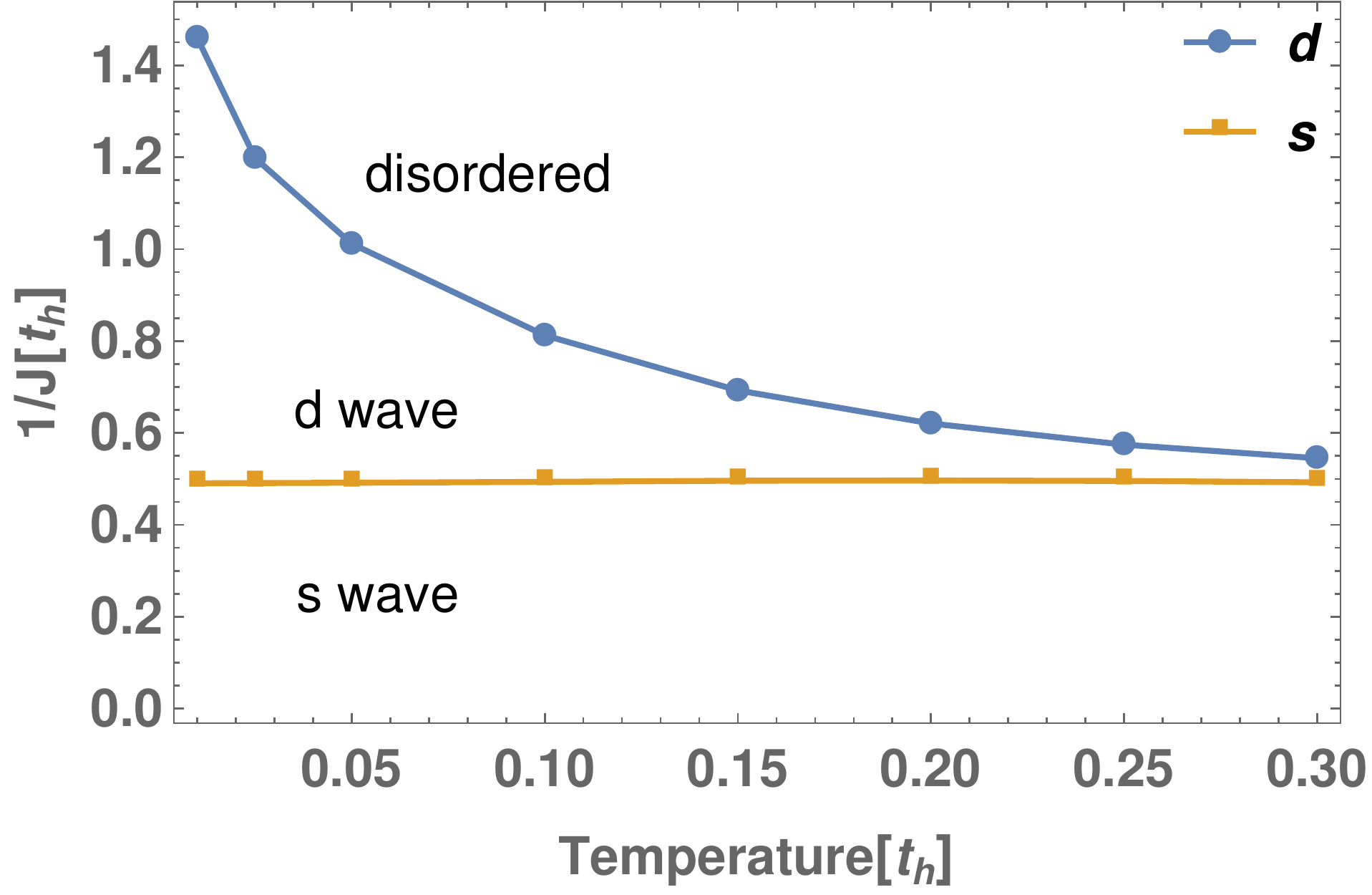}
\caption{
Graphene, doping $\delta=0.1$. The phase diagram for graphene determined by the fastest growing order parameter.
The line corresponding to the transition from the disordered (normal) phase to the $d$-wave phase coincides with the equilibrium phase diagram.
As the quench amplitude is increased, the $s$-wave phase is preferred.
}
\label{fig3}
\end{figure}

\section{Results} \label{results}

We now present results for the solution of the evolution equation Eq.~\eqref{eomDel2}. We start with an initial small (precise number does not matter),
random vector $\vec{\Delta}$ at $t=0$, and time-evolve it forward. We project the time-evolution along the three mutually orthogonal eigenvectors of the 
pairing susceptibility $\Pi^R$.
For graphene these correspond to the $s$-wave order parameter and a doubly degenerate sub-space with $d$-wave symmetry. For the Haldane model,
the degeneracy is lifted into two chiral solutions $d+id$, and $d-id$ respectively.

The typical time-evolution of $\vec{\Delta}$ for the Haldane model is shown in Fig.~\ref{fig1}. The slopes are proportional to the growth rate,
and for the parameters chosen, $d+id$ grows faster than $d-id$, followed by $s$-wave. The magnitude of the order-parameter at any given time depends 
on the initial condition, and can be rescaled away, and is therefore not of physical relevance. Only the growth rates convey the main physics.
For this case we conclude that the
preferred superconducting phase is $d+id$. 

The time-evolution for doped graphene is similar to
Fig.~\ref{fig1} except that the slopes for $d+id$ and $d-id$
are the same, reflecting the degenerate eigenvalues of $\Pi^R$. While Fig.~\ref{fig1} is for parameters where
all three orders grow in time, the disordered phase is characterized by all three orders decaying exponentially in time.
In addition, for $J$-values smaller than that shown in Fig.~\ref{fig1}, we can have a situation,
where only one order-parameter (typically $d$-wave for the doping levels discussed) grows,
while the others decay in time. For all these scenarios, the fastest growing order-parameter determines the preferred phase of
the system.

\begin{figure}
\includegraphics[width = 1.0\columnwidth]{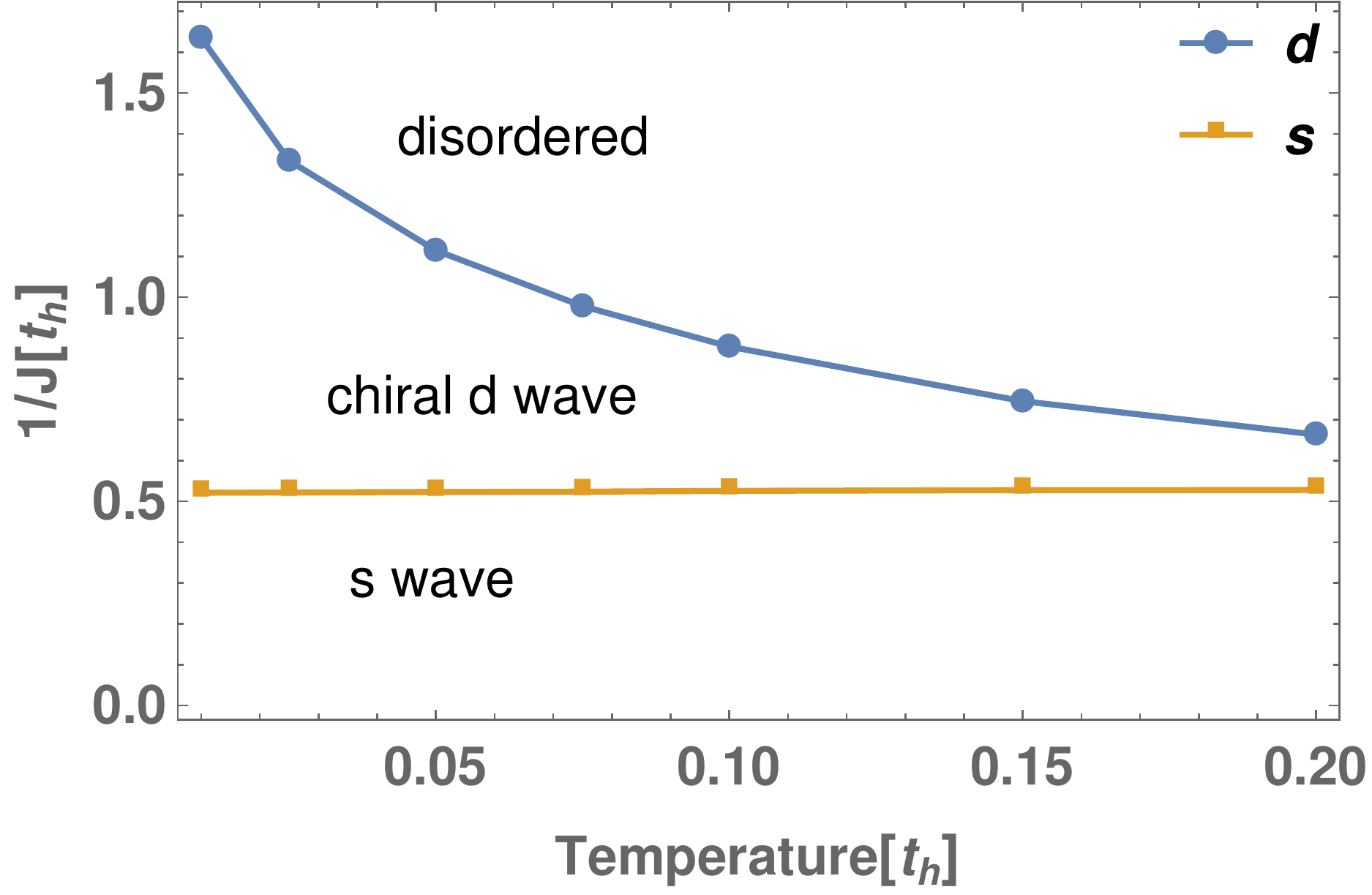}
\caption{
Haldane model $A_0a=0.5,\Omega=10t_h$ and doping $\delta=0.1$. The phase diagram determined by the fastest growing order parameter.
The line corresponding to the transition from the disordered (normal) phase to the chiral $d$-wave phase coincides with the equilibrium phase diagram.
As the quench amplitude is increased, the $s$-wave phase is preferred.
}
\label{fig4}
\end{figure}

As the quench amplitude increases, the $d$-wave order parameter growth rate increases. However, after some critical value
of $J$, the $s$-wave order parameter begins to grow faster, eventually out growing the $d$-wave order parameter. How the growth
rate difference of the $d$-wave and $s$-wave order parameters vary with quench amplitude is shown in Fig.~\ref{fig2}.
Initially the $d$-wave becomes more unstable as the interaction parameter increases. This result is expected as in general
the growth rate of any order parameter is zero at the critical point, and grows (decreases) faster as one moves further into (away) from
the ordered phase. However, we find that after a critical $J$,
the difference between the growth rates of the two order parameters rapidly approaches zero, with the $s$-wave growing faster than
the $d$-wave beyond a critical quench amplitude.
\begin{figure}
\includegraphics[width = 1.0\columnwidth]{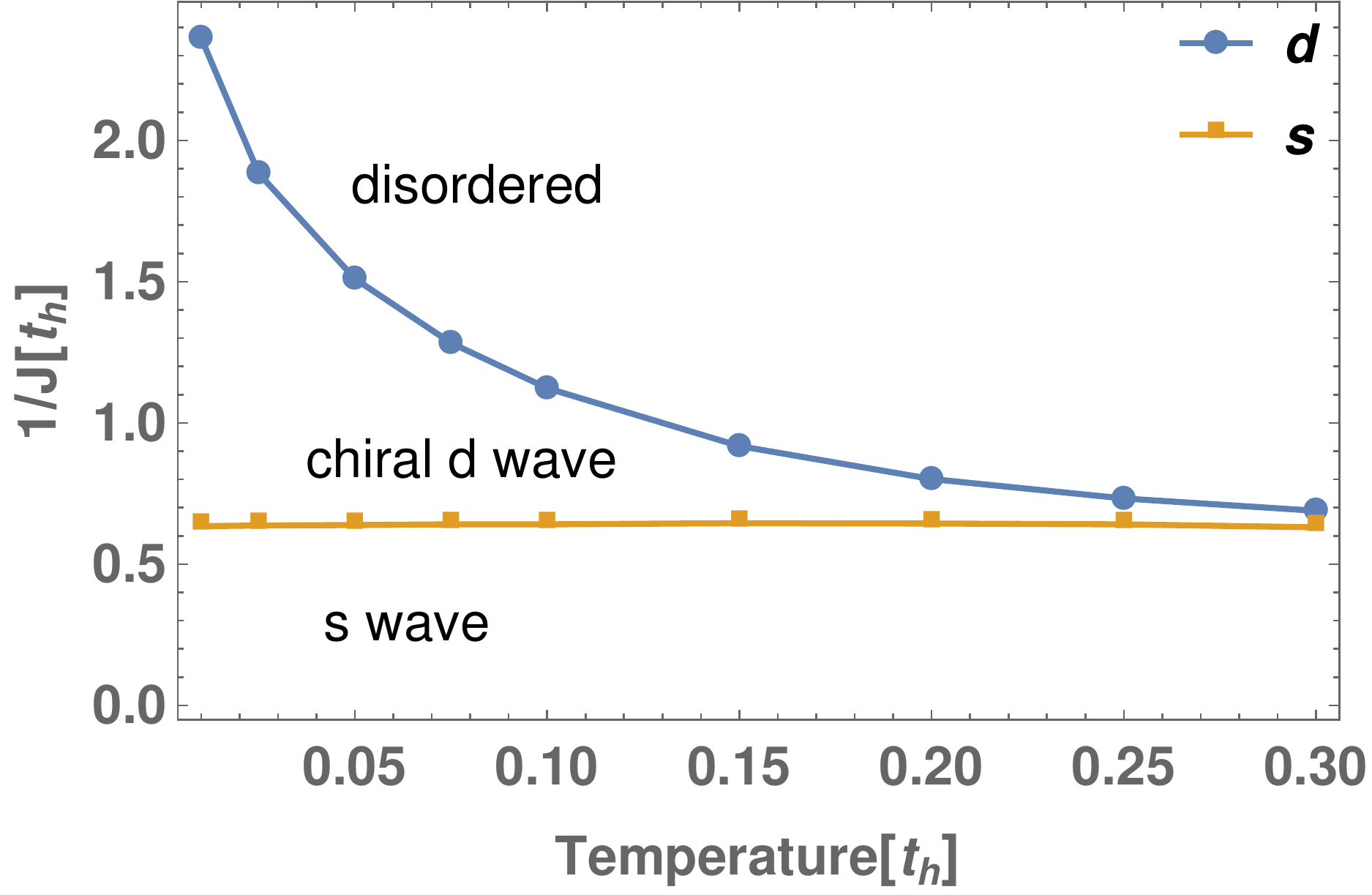}
\caption{
Haldane model $A_0a=1.0,\Omega=10t_h$ and doping $\delta=0.1$. The phase diagram determined by the fastest growing order parameter.
The line corresponding to the transition from the disordered (normal) phase to the chiral $d$-wave phase coincides with the equilibrium phase diagram.
As the quench amplitude is increased, the $s$-wave phase is preferred.
}
\label{fig5}
\end{figure}

The phase diagram determined by the fastest growing order parameter is shown in Fig.~\ref{fig3} for doped graphene,
and Figures~\ref{fig4} and~\ref{fig5} for the Haldane model. The two figures for the Haldane model are obtained by applying a laser of 
two different strengths but same frequency to graphene. We still take care that the frequency of the laser is larger than the bandwidth
of graphene so that the high-frequency expansion is valid.
 
For all these figures, the transition from the
disordered (i.e, normal) to the $d$-wave ordered phase obtained from the time-evolution, coincides with an equilibrium 
calculation based on identifying the $T$ and $J$ values at which $1= J\Pi^R(\omega=0)$. 
Thus consistent with BCS theory, the boundary between the disordered and $d$-wave phase in Figures~\ref{fig3},~\ref{fig4},~\ref{fig5} 
are strongly dependent on the density of states at the Fermi energy. For this reason, the onset of superconductivity in
Fig.~\ref{fig5} occurs at a smaller value of the interaction as compared to the other figures because
the larger amplitude laser in Fig.~\ref{fig5} flattens the bands more, increasing the density of states at the Fermi energy.

For larger values of $J$ (i.e., as the quench
amplitude increases), eventually the $s$-wave grows faster for graphene as well as the Haldane model.
This result can be understood as follows. For smaller quench amplitudes, the dynamics is
primarily from quasi-particle excitations in the vicinity of the Fermi surface. Thus the equilibrium phase
predicting $d$-wave order is recovered. In contrast, for larger quench amplitudes,
the dynamics is governed by highly excited quasi-particles. These quasi-particles cause an effective dephasing, leading
to an averaging over the entire Brillouin zone (BZ). This averaging favors an $s$-wave rather than a $d$-wave because the latter order parameter
changes sign in the BZ, so that its magnitude is effectively averaged out to zero by the dephasing.

Since at short times $t\sim O(1)$, quasi-particles everywhere in the BZ participate in the dynamics,
the $s$-wave component of the order-parameter grows
faster than the $d$-wave (see Fig.~\ref{fig1}). If the quench amplitude is large, the initial impulse on the $s$-wave order-parameter
is large enough to overtake the growth of the $d$-wave. For the particular case shown in Fig.~\ref{fig1}, this initial impulse is not strong enough,
and at long times the $d$-wave grows faster.

Note that the boundary between the $d$-wave and $s$-wave phases in Figures~\ref{fig3},~\ref{fig4} and~\ref{fig5}
is rather flat as a function of temperature.
This is simply reflecting the fact that since quasi-particles everywhere in the BZ are participating in the dynamics for such large
quench amplitudes, these are not
sensitive to the details of the initial distribution function. The location of the flat line between
the $d$-wave and $s$-wave phases does however depend on the initial state through the doping level.
\begin{figure}
\includegraphics[width = 1.0\columnwidth]{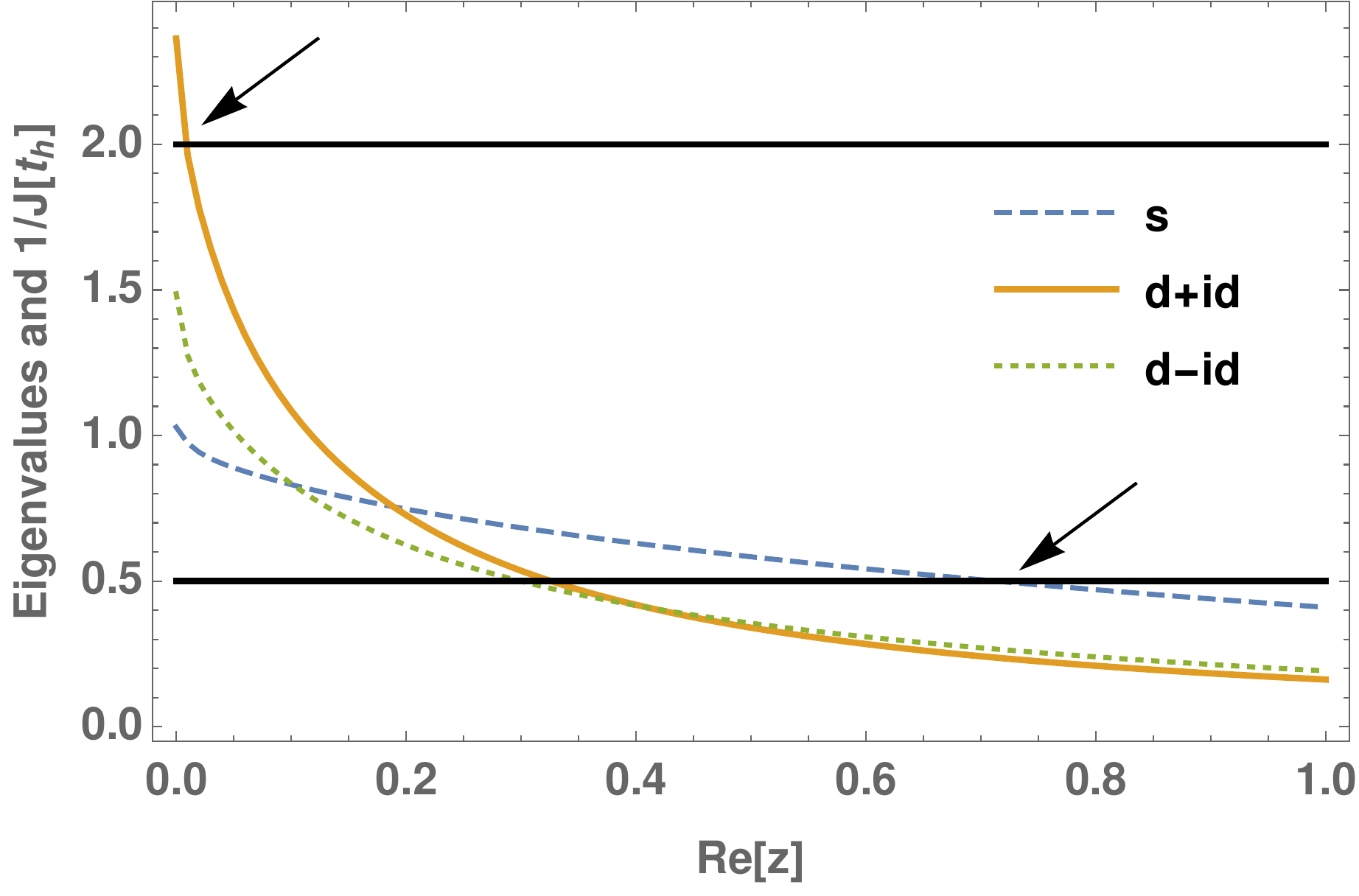}
\caption{
Haldane model $A_0a=1.0,\Omega=10t_h,T=0.01t_h,\delta=0.1$. Plot of eigenvalues of ${\rm Re}\biggl[\Pi^R(z)\biggr]$.
Two horizontal solid lines indicate two different values of the
inverse coupling $1/J$.
The rate of growth of the order parameter is approximately $2 z_0$ where $z_0$ is the zeros of the function $1/J-\Pi^R(z_0)=0$. The zeros for the two
different couplings are indicated by the two arrows. The fastest growing
order parameter changes from chiral $d$-wave (top-left arrow) to $s$-wave (bottom-right arrow) as the quench amplitude is increased.
}
\label{fig6}
\end{figure}

In all the three Figures~\ref{fig3},~\ref{fig4} and~\ref{fig5}, we plot the phase diagram up to the temperatures where the
$d$-wave and the $s$-wave growth rates are clearly different. At higher temperatures, all the order parameter growth rates
are degenerate, and one has to account for effects beyond mean-field in order to lift their degeneracy,
an analysis that is beyond the scope of the paper.

We make our observation regarding the favored superconducting order more formal by noting that the growth or decay rate of an
initial order parameter fluctuation is determined by the location
of the zeros of $D(q,z)=1-J\Pi^R(q,z)$ in the complex $z$ plane (see Eq.~\eqref{Dz} and discussion above it).
To locate the zero $z_0$, the real and imaginary parts of $z$ should be tuned
together to satisfy the two conditions $1/J = {\rm Re}[\Pi^R(z_0)]=0, {\rm Im}[\Pi^R(z_0)]=0$. The real part of the zero ${\rm Re}\left[z_0\right]$
determines the exponential growth or decay rate, while the imaginary part ${\rm Im}\left[z_0\right]$ denotes an overall oscillation.

In equilibrium, the critical temperature $T_c$ for
a given $J$ is one where the zero approaches $z=0$ linearly in the deviation from the critical temperature or critical coupling.
For an interaction quench deeper and deeper into the ordered phase, the pole shifts further out into
the complex plane, indicating that the order parameter grows faster. This behavior can be seen in Fig.~\ref{fig6} where the growth rate is determined
by the value of $z$ for which $1/J$ intersects with the real part of $\Pi^R(z)$. In addition, Fig.~\ref{fig6} clearly shows that as $J$ increases,
first the $d$-wave grows faster. But for larger values of $J$, eventually the $s$-wave order grows fastest.

In Fig.~\ref{fig6}, for simplicity we allow $z$ to be purely real, and so we assume that the zeros fall
entirely on the imaginary axis, or equivalently the
order-parameter has a purely exponentially growing or decaying component in time,
and no oscillatory component. We look for solutions where $1/J={\rm Re}[\Pi^R(z)]$. Inspite of not doing the analysis in the full complex plane,
the growth rate and critical couplings agree very well with the full time evolution. Since the full correct analysis is the real time evolution,
this good agreement between the time-evolution and the zeros in the complex plane indicate that ${\rm Im}\left[z_0\right]\simeq 0$.

We now discuss the effect of inelastic scattering. This will have two effects. One is to thermalize the electron gas,
giving rise to a temperature larger than the initial temperature $T$ of the electron gas. As long as the final temperature
is less than $T_{\rm BKT}$, the system will show quasi-long range order at long times.
The second effect of inelastic
scattering is that it causes the breakdown of the mean-field approximation by causing scattering between
superconducting fluctuations. 
For the parameters of Fig.~\ref{fig1}, the Fermi Golden rule
estimate for the inelastic scattering time (which is also the time for leaving the prethermal regime) is $t_{\rm in}=1/J^2T$ $\sim 30 t_h^{-1}$.  
Fig.~\ref{fig1} shows that the physical
quantity of interest, the growth rate or slope has stabilized well before this time. In addition the growth rate itself is 
order of the hopping amplitude, $\nu_{s,d\pm id} \sim t_h$, and is therefore much
larger than the inelastic scattering rate ($t_{\rm in}^{-1}\sim 0.03 t_h$). Thus for the parameters discussed in the paper, the transient regime is indeed dominated by the
mean field estimate because the scattering rate between fluctuations is small as compared to the rate at which the order-parameter grows.

\section{Conclusions} \label{concl}

The growth of superconducting order under nonequilibrium conditions is an active area of research.
In this paper we give an explicit example of how two competing superconducting states behave
after an interaction quench. We show that tuning the interaction quench amplitude
can favor one superconducting state over the other.

This result is relevant to experiments
in cold-atom gases where interaction quenches into the superconducting phase can be performed.
Recent experiments in pump-probe spectroscopy make these results also relevant to the solid-state.
In fact a laser can modify the lattice  parameters, and through it the effective interactions in the
$t$-$J$ model used by us to study superconductivity. Thus even in the solid-state, a laser quench can
effectively give rise to an interaction quench.

While our study was for a hexagonal lattice with and without TR symmetry, our result showing how the dominant
phase can be tuned by nonequilibrium conditions, such as an external laser and interaction quenches, is rather general
and applicable to a wide range of lattice models where different ordered phases compete.

Our study treated the effect of the TR symmetry breaking laser within a high frequency expansion.
The dynamics in the presence of a resonant low-frequency laser is expected to be even
richer, and is left for future studies. Another interesting and challenging direction of study is to go beyond mean-field
by allowing for interactions between the superconducting fluctuations. Since a laser, by its spatial orientation
at any given time, breaks the underlying
lattice symmetries,
the interactions in principle can couple different superconducting channels. This could affect the outcome
of the symmetry of the dominant superconducting order-parameter at steady-state.

{\sl Acknowledgements:}
The authors thank Yonah Lemonik for many useful discussions and Daniel Yates for
numerical help.
This work was supported by the US Department of Energy,
Office of Science, Basic Energy Sciences, under Award No.~DE-SC0010821.


\begin{thebibliography}{43}%
\makeatletter
\providecommand \@ifxundefined [1]{%
 \@ifx{#1\undefined}
}%
\providecommand \@ifnum [1]{%
 \ifnum #1\expandafter \@firstoftwo
 \else \expandafter \@secondoftwo
 \fi
}%
\providecommand \@ifx [1]{%
 \ifx #1\expandafter \@firstoftwo
 \else \expandafter \@secondoftwo
 \fi
}%
\providecommand \natexlab [1]{#1}%
\providecommand \enquote  [1]{``#1''}%
\providecommand \bibnamefont  [1]{#1}%
\providecommand \bibfnamefont [1]{#1}%
\providecommand \citenamefont [1]{#1}%
\providecommand \href@noop [0]{\@secondoftwo}%
\providecommand \href [0]{\begingroup \@sanitize@url \@href}%
\providecommand \@href[1]{\@@startlink{#1}\@@href}%
\providecommand \@@href[1]{\endgroup#1\@@endlink}%
\providecommand \@sanitize@url [0]{\catcode `\\12\catcode `\$12\catcode
  `\&12\catcode `\#12\catcode `\^12\catcode `\_12\catcode `\%12\relax}%
\providecommand \@@startlink[1]{}%
\providecommand \@@endlink[0]{}%
\providecommand \url  [0]{\begingroup\@sanitize@url \@url }%
\providecommand \@url [1]{\endgroup\@href {#1}{\urlprefix }}%
\providecommand \urlprefix  [0]{URL }%
\providecommand \Eprint [0]{\href }%
\providecommand \doibase [0]{http://dx.doi.org/}%
\providecommand \selectlanguage [0]{\@gobble}%
\providecommand \bibinfo  [0]{\@secondoftwo}%
\providecommand \bibfield  [0]{\@secondoftwo}%
\providecommand \translation [1]{[#1]}%
\providecommand \BibitemOpen [0]{}%
\providecommand \bibitemStop [0]{}%
\providecommand \bibitemNoStop [0]{.\EOS\space}%
\providecommand \EOS [0]{\spacefactor3000\relax}%
\providecommand \BibitemShut  [1]{\csname bibitem#1\endcsname}%
\let\auto@bib@innerbib\@empty
\bibitem [{\citenamefont {Fausti}\ \emph {et~al.}(2011)\citenamefont {Fausti},
  \citenamefont {Tobey}, \citenamefont {Dean}, \citenamefont {Kaiser},
  \citenamefont {Dienst}, \citenamefont {Hoffmann}, \citenamefont {Pyon},
  \citenamefont {Takayama}, \citenamefont {Takagi},\ and\ \citenamefont
  {Cavalleri}}]{Fausti11}%
  \BibitemOpen
  \bibfield  {author} {\bibinfo {author} {\bibfnamefont {D.}~\bibnamefont
  {Fausti}}, \bibinfo {author} {\bibfnamefont {R.~I.}\ \bibnamefont {Tobey}},
  \bibinfo {author} {\bibfnamefont {N.}~\bibnamefont {Dean}}, \bibinfo {author}
  {\bibfnamefont {S.}~\bibnamefont {Kaiser}}, \bibinfo {author} {\bibfnamefont
  {A.}~\bibnamefont {Dienst}}, \bibinfo {author} {\bibfnamefont {M.~C.}\
  \bibnamefont {Hoffmann}}, \bibinfo {author} {\bibfnamefont {S.}~\bibnamefont
  {Pyon}}, \bibinfo {author} {\bibfnamefont {T.}~\bibnamefont {Takayama}},
  \bibinfo {author} {\bibfnamefont {H.}~\bibnamefont {Takagi}}, \ and\ \bibinfo
  {author} {\bibfnamefont {A.}~\bibnamefont {Cavalleri}},\ }\href {\doibase
  10.1126/science.1197294} {\bibfield  {journal} {\bibinfo  {journal}
  {Science}\ }\textbf {\bibinfo {volume} {331}},\ \bibinfo {pages} {189}
  (\bibinfo {year} {2011})}\BibitemShut {NoStop}%
\bibitem [{\citenamefont {Graf}\ \emph {et~al.}(2011)\citenamefont {Graf},
  \citenamefont {Jozwiak}, \citenamefont {Smallwood}, \citenamefont {Eisaki},
  \citenamefont {Kaindl}, \citenamefont {Lee},\ and\ \citenamefont
  {Lanzara}}]{Graf11}%
  \BibitemOpen
  \bibfield  {author} {\bibinfo {author} {\bibfnamefont {J.}~\bibnamefont
  {Graf}}, \bibinfo {author} {\bibfnamefont {C.}~\bibnamefont {Jozwiak}},
  \bibinfo {author} {\bibfnamefont {C.~L.}\ \bibnamefont {Smallwood}}, \bibinfo
  {author} {\bibfnamefont {H.}~\bibnamefont {Eisaki}}, \bibinfo {author}
  {\bibfnamefont {R.~A.}\ \bibnamefont {Kaindl}}, \bibinfo {author}
  {\bibfnamefont {D.-H.}\ \bibnamefont {Lee}}, \ and\ \bibinfo {author}
  {\bibfnamefont {A.}~\bibnamefont {Lanzara}},\ }\href@noop {} {\bibfield
  {journal} {\bibinfo  {journal} {Nature Physics}\ }\textbf {\bibinfo {volume}
  {7}},\ \bibinfo {pages} {805} (\bibinfo {year} {2011})}\BibitemShut {NoStop}%
\bibitem [{\citenamefont {Smallwood}\ \emph {et~al.}(2012)\citenamefont
  {Smallwood}, \citenamefont {Hinton}, \citenamefont {Jozwiak}, \citenamefont
  {Zhang}, \citenamefont {Koralek}, \citenamefont {Eisaki}, \citenamefont
  {Lee}, \citenamefont {Orenstein},\ and\ \citenamefont
  {Lanzara}}]{Smallwood12}%
  \BibitemOpen
  \bibfield  {author} {\bibinfo {author} {\bibfnamefont {C.~L.}\ \bibnamefont
  {Smallwood}}, \bibinfo {author} {\bibfnamefont {J.~P.}\ \bibnamefont
  {Hinton}}, \bibinfo {author} {\bibfnamefont {C.}~\bibnamefont {Jozwiak}},
  \bibinfo {author} {\bibfnamefont {W.}~\bibnamefont {Zhang}}, \bibinfo
  {author} {\bibfnamefont {J.~D.}\ \bibnamefont {Koralek}}, \bibinfo {author}
  {\bibfnamefont {H.}~\bibnamefont {Eisaki}}, \bibinfo {author} {\bibfnamefont
  {D.-H.}\ \bibnamefont {Lee}}, \bibinfo {author} {\bibfnamefont
  {J.}~\bibnamefont {Orenstein}}, \ and\ \bibinfo {author} {\bibfnamefont
  {A.}~\bibnamefont {Lanzara}},\ }\href {\doibase 10.1126/science.1217423}
  {\bibfield  {journal} {\bibinfo  {journal} {Science}\ }\textbf {\bibinfo
  {volume} {336}},\ \bibinfo {pages} {1137} (\bibinfo {year}
  {2012})}\BibitemShut {NoStop}%
\bibitem [{\citenamefont {Smallwood}\ \emph {et~al.}(2014)\citenamefont
  {Smallwood}, \citenamefont {Zhang}, \citenamefont {Miller}, \citenamefont
  {Jozwiak}, \citenamefont {Eisaki}, \citenamefont {Lee},\ and\ \citenamefont
  {Lanzara}}]{Smallwood14}%
  \BibitemOpen
  \bibfield  {author} {\bibinfo {author} {\bibfnamefont {C.~L.}\ \bibnamefont
  {Smallwood}}, \bibinfo {author} {\bibfnamefont {W.}~\bibnamefont {Zhang}},
  \bibinfo {author} {\bibfnamefont {T.~L.}\ \bibnamefont {Miller}}, \bibinfo
  {author} {\bibfnamefont {C.}~\bibnamefont {Jozwiak}}, \bibinfo {author}
  {\bibfnamefont {H.}~\bibnamefont {Eisaki}}, \bibinfo {author} {\bibfnamefont
  {D.-H.}\ \bibnamefont {Lee}}, \ and\ \bibinfo {author} {\bibfnamefont
  {A.}~\bibnamefont {Lanzara}},\ }\href {\doibase 10.1103/PhysRevB.89.115126}
  {\bibfield  {journal} {\bibinfo  {journal} {Phys. Rev. B}\ }\textbf {\bibinfo
  {volume} {89}},\ \bibinfo {pages} {115126} (\bibinfo {year}
  {2014})}\BibitemShut {NoStop}%
\bibitem [{\citenamefont {Beck}\ \emph {et~al.}(2013)\citenamefont {Beck},
  \citenamefont {Rousseau}, \citenamefont {Klammer}, \citenamefont {Leiderer},
  \citenamefont {Mittendorff}, \citenamefont {Winnerl}, \citenamefont {Helm},
  \citenamefont {Gol'tsman},\ and\ \citenamefont {Demsar}}]{Beck13}%
  \BibitemOpen
  \bibfield  {author} {\bibinfo {author} {\bibfnamefont {M.}~\bibnamefont
  {Beck}}, \bibinfo {author} {\bibfnamefont {I.}~\bibnamefont {Rousseau}},
  \bibinfo {author} {\bibfnamefont {M.}~\bibnamefont {Klammer}}, \bibinfo
  {author} {\bibfnamefont {P.}~\bibnamefont {Leiderer}}, \bibinfo {author}
  {\bibfnamefont {M.}~\bibnamefont {Mittendorff}}, \bibinfo {author}
  {\bibfnamefont {S.}~\bibnamefont {Winnerl}}, \bibinfo {author} {\bibfnamefont
  {M.}~\bibnamefont {Helm}}, \bibinfo {author} {\bibfnamefont {G.~N.}\
  \bibnamefont {Gol'tsman}}, \ and\ \bibinfo {author} {\bibfnamefont
  {J.}~\bibnamefont {Demsar}},\ }\href {\doibase
  10.1103/PhysRevLett.110.267003} {\bibfield  {journal} {\bibinfo  {journal}
  {Phys. Rev. Lett.}\ }\textbf {\bibinfo {volume} {110}},\ \bibinfo {pages}
  {267003} (\bibinfo {year} {2013})}\BibitemShut {NoStop}%
\bibitem [{\citenamefont {Mitrano}\ \emph {et~al.}(2016)\citenamefont
  {Mitrano}, \citenamefont {Cantaluppi}, \citenamefont {Nicoletti},
  \citenamefont {Kaiser}, \citenamefont {Perucchi}, \citenamefont {Lupi},
  \citenamefont {Pietro}, \citenamefont {Pontiroli}, \citenamefont {Ricc\'{o}},
  \citenamefont {Clark}, \citenamefont {Jaksch},\ and\ \citenamefont
  {Cavalleri}}]{Mitrano15}%
  \BibitemOpen
  \bibfield  {author} {\bibinfo {author} {\bibfnamefont {M.}~\bibnamefont
  {Mitrano}}, \bibinfo {author} {\bibfnamefont {A.}~\bibnamefont {Cantaluppi}},
  \bibinfo {author} {\bibfnamefont {D.}~\bibnamefont {Nicoletti}}, \bibinfo
  {author} {\bibfnamefont {S.}~\bibnamefont {Kaiser}}, \bibinfo {author}
  {\bibfnamefont {A.}~\bibnamefont {Perucchi}}, \bibinfo {author}
  {\bibfnamefont {S.}~\bibnamefont {Lupi}}, \bibinfo {author} {\bibfnamefont
  {P.~D.}\ \bibnamefont {Pietro}}, \bibinfo {author} {\bibfnamefont
  {D.}~\bibnamefont {Pontiroli}}, \bibinfo {author} {\bibfnamefont
  {M.}~\bibnamefont {Ricc\'{o}}}, \bibinfo {author} {\bibfnamefont {S.~R.}\
  \bibnamefont {Clark}}, \bibinfo {author} {\bibfnamefont {D.}~\bibnamefont
  {Jaksch}}, \ and\ \bibinfo {author} {\bibfnamefont {A.}~\bibnamefont
  {Cavalleri}},\ }\href@noop {} {\bibfield  {journal} {\bibinfo  {journal}
  {Nature}\ }\textbf {\bibinfo {volume} {530}},\ \bibinfo {pages} {461}
  (\bibinfo {year} {2016})}\BibitemShut {NoStop}%
\bibitem [{\citenamefont {Regal}\ \emph {et~al.}(2004)\citenamefont {Regal},
  \citenamefont {Greiner},\ and\ \citenamefont {Jin}}]{Regal04}%
  \BibitemOpen
  \bibfield  {author} {\bibinfo {author} {\bibfnamefont {C.~A.}\ \bibnamefont
  {Regal}}, \bibinfo {author} {\bibfnamefont {M.}~\bibnamefont {Greiner}}, \
  and\ \bibinfo {author} {\bibfnamefont {D.~S.}\ \bibnamefont {Jin}},\ }\href
  {\doibase 10.1103/PhysRevLett.92.040403} {\bibfield  {journal} {\bibinfo
  {journal} {Phys. Rev. Lett.}\ }\textbf {\bibinfo {volume} {92}},\ \bibinfo
  {pages} {040403} (\bibinfo {year} {2004})}\BibitemShut {NoStop}%
\bibitem [{\citenamefont {Zwierlein}\ \emph {et~al.}(2004)\citenamefont
  {Zwierlein}, \citenamefont {Stan}, \citenamefont {Schunck}, \citenamefont
  {Raupach}, \citenamefont {Kerman},\ and\ \citenamefont
  {Ketterle}}]{Zwierlein04}%
  \BibitemOpen
  \bibfield  {author} {\bibinfo {author} {\bibfnamefont {M.~W.}\ \bibnamefont
  {Zwierlein}}, \bibinfo {author} {\bibfnamefont {C.~A.}\ \bibnamefont {Stan}},
  \bibinfo {author} {\bibfnamefont {C.~H.}\ \bibnamefont {Schunck}}, \bibinfo
  {author} {\bibfnamefont {S.~M.~F.}\ \bibnamefont {Raupach}}, \bibinfo
  {author} {\bibfnamefont {A.~J.}\ \bibnamefont {Kerman}}, \ and\ \bibinfo
  {author} {\bibfnamefont {W.}~\bibnamefont {Ketterle}},\ }\href {\doibase
  10.1103/PhysRevLett.92.120403} {\bibfield  {journal} {\bibinfo  {journal}
  {Phys. Rev. Lett.}\ }\textbf {\bibinfo {volume} {92}},\ \bibinfo {pages}
  {120403} (\bibinfo {year} {2004})}\BibitemShut {NoStop}%
\bibitem [{\citenamefont {Bloch}\ \emph {et~al.}(2008)\citenamefont {Bloch},
  \citenamefont {Dalibard},\ and\ \citenamefont {Zwerger}}]{Bloch08}%
  \BibitemOpen
  \bibfield  {author} {\bibinfo {author} {\bibfnamefont {I.}~\bibnamefont
  {Bloch}}, \bibinfo {author} {\bibfnamefont {J.}~\bibnamefont {Dalibard}}, \
  and\ \bibinfo {author} {\bibfnamefont {W.}~\bibnamefont {Zwerger}},\ }\href
  {\doibase 10.1103/RevModPhys.80.885} {\bibfield  {journal} {\bibinfo
  {journal} {Rev. Mod. Phys.}\ }\textbf {\bibinfo {volume} {80}},\ \bibinfo
  {pages} {885} (\bibinfo {year} {2008})}\BibitemShut {NoStop}%
\bibitem [{\citenamefont {Endres}\ \emph {et~al.}(2012)\citenamefont {Endres},
  \citenamefont {Fukuhara}, \citenamefont {Pekker}, , \citenamefont {Cheneau},
  \citenamefont {Schaub}, \citenamefont {Gross}, \citenamefont {Demler},
  \citenamefont {Kuhr},\ and\ \citenamefont {Bloch}}]{Bloch12}%
  \BibitemOpen
  \bibfield  {author} {\bibinfo {author} {\bibfnamefont {M.}~\bibnamefont
  {Endres}}, \bibinfo {author} {\bibfnamefont {T.}~\bibnamefont {Fukuhara}},
  \bibinfo {author} {\bibfnamefont {D.}~\bibnamefont {Pekker}}, , \bibinfo
  {author} {\bibfnamefont {M.}~\bibnamefont {Cheneau}}, \bibinfo {author}
  {\bibfnamefont {P.}~\bibnamefont {Schaub}}, \bibinfo {author} {\bibfnamefont
  {C.}~\bibnamefont {Gross}}, \bibinfo {author} {\bibfnamefont
  {E.}~\bibnamefont {Demler}}, \bibinfo {author} {\bibfnamefont
  {S.}~\bibnamefont {Kuhr}}, \ and\ \bibinfo {author} {\bibfnamefont
  {I.}~\bibnamefont {Bloch}},\ }\href@noop {} {\bibfield  {journal} {\bibinfo
  {journal} {Nature}\ }\textbf {\bibinfo {volume} {487}},\ \bibinfo {pages}
  {454} (\bibinfo {year} {2012})}\BibitemShut {NoStop}%
\bibitem [{\citenamefont {Yin}\ and\ \citenamefont
  {Radzihovsky}(2016)}]{Yin16}%
  \BibitemOpen
  \bibfield  {author} {\bibinfo {author} {\bibfnamefont {X.}~\bibnamefont
  {Yin}}\ and\ \bibinfo {author} {\bibfnamefont {L.}~\bibnamefont
  {Radzihovsky}},\ }\href {\doibase 10.1103/PhysRevA.93.033653} {\bibfield
  {journal} {\bibinfo  {journal} {Phys. Rev. A}\ }\textbf {\bibinfo {volume}
  {93}},\ \bibinfo {pages} {033653} (\bibinfo {year} {2016})}\BibitemShut
  {NoStop}%
\bibitem [{\citenamefont {Yuzbashyan}\ \emph {et~al.}(2015)\citenamefont
  {Yuzbashyan}, \citenamefont {Dzero}, \citenamefont {Gurarie},\ and\
  \citenamefont {Foster}}]{Yuzbashyan15}%
  \BibitemOpen
  \bibfield  {author} {\bibinfo {author} {\bibfnamefont {E.~A.}\ \bibnamefont
  {Yuzbashyan}}, \bibinfo {author} {\bibfnamefont {M.}~\bibnamefont {Dzero}},
  \bibinfo {author} {\bibfnamefont {V.}~\bibnamefont {Gurarie}}, \ and\
  \bibinfo {author} {\bibfnamefont {M.~S.}\ \bibnamefont {Foster}},\ }\href
  {\doibase 10.1103/PhysRevA.91.033628} {\bibfield  {journal} {\bibinfo
  {journal} {Phys. Rev. A}\ }\textbf {\bibinfo {volume} {91}},\ \bibinfo
  {pages} {033628} (\bibinfo {year} {2015})}\BibitemShut {NoStop}%
\bibitem [{\citenamefont {Liao}\ and\ \citenamefont {Foster}(2015)}]{Foster15}%
  \BibitemOpen
  \bibfield  {author} {\bibinfo {author} {\bibfnamefont {Y.}~\bibnamefont
  {Liao}}\ and\ \bibinfo {author} {\bibfnamefont {M.~S.}\ \bibnamefont
  {Foster}},\ }\href {\doibase 10.1103/PhysRevA.92.053620} {\bibfield
  {journal} {\bibinfo  {journal} {Phys. Rev. A}\ }\textbf {\bibinfo {volume}
  {92}},\ \bibinfo {pages} {053620} (\bibinfo {year} {2015})}\BibitemShut
  {NoStop}%
\bibitem [{\citenamefont {Foster}\ \emph {et~al.}(2014)\citenamefont {Foster},
  \citenamefont {Gurarie}, \citenamefont {Dzero},\ and\ \citenamefont
  {Yuzbashyan}}]{Foster14}%
  \BibitemOpen
  \bibfield  {author} {\bibinfo {author} {\bibfnamefont {M.~S.}\ \bibnamefont
  {Foster}}, \bibinfo {author} {\bibfnamefont {V.}~\bibnamefont {Gurarie}},
  \bibinfo {author} {\bibfnamefont {M.}~\bibnamefont {Dzero}}, \ and\ \bibinfo
  {author} {\bibfnamefont {E.~A.}\ \bibnamefont {Yuzbashyan}},\ }\href
  {\doibase 10.1103/PhysRevLett.113.076403} {\bibfield  {journal} {\bibinfo
  {journal} {Phys. Rev. Lett.}\ }\textbf {\bibinfo {volume} {113}},\ \bibinfo
  {pages} {076403} (\bibinfo {year} {2014})}\BibitemShut {NoStop}%
\bibitem [{\citenamefont {Sentef}\ \emph {et~al.}(2016)\citenamefont {Sentef},
  \citenamefont {Kemper}, \citenamefont {Georges},\ and\ \citenamefont
  {Kollath}}]{Sentef16}%
  \BibitemOpen
  \bibfield  {author} {\bibinfo {author} {\bibfnamefont {M.~A.}\ \bibnamefont
  {Sentef}}, \bibinfo {author} {\bibfnamefont {A.~F.}\ \bibnamefont {Kemper}},
  \bibinfo {author} {\bibfnamefont {A.}~\bibnamefont {Georges}}, \ and\
  \bibinfo {author} {\bibfnamefont {C.}~\bibnamefont {Kollath}},\ }\href
  {\doibase 10.1103/PhysRevB.93.144506} {\bibfield  {journal} {\bibinfo
  {journal} {Phys. Rev. B}\ }\textbf {\bibinfo {volume} {93}},\ \bibinfo
  {pages} {144506} (\bibinfo {year} {2016})}\BibitemShut {NoStop}%
\bibitem [{\citenamefont {Knap}\ \emph {et~al.}(2016)\citenamefont {Knap},
  \citenamefont {Babadi}, \citenamefont {Refael}, \citenamefont {Martin},\ and\
  \citenamefont {Demler}}]{Knap16}%
  \BibitemOpen
  \bibfield  {author} {\bibinfo {author} {\bibfnamefont {M.}~\bibnamefont
  {Knap}}, \bibinfo {author} {\bibfnamefont {M.}~\bibnamefont {Babadi}},
  \bibinfo {author} {\bibfnamefont {G.}~\bibnamefont {Refael}}, \bibinfo
  {author} {\bibfnamefont {I.}~\bibnamefont {Martin}}, \ and\ \bibinfo {author}
  {\bibfnamefont {E.}~\bibnamefont {Demler}},\ }\href {\doibase
  10.1103/PhysRevB.94.214504} {\bibfield  {journal} {\bibinfo  {journal} {Phys.
  Rev. B}\ }\textbf {\bibinfo {volume} {94}},\ \bibinfo {pages} {214504}
  (\bibinfo {year} {2016})}\BibitemShut {NoStop}%
\bibitem [{\citenamefont {Kennes}\ \emph {et~al.}(2017)\citenamefont {Kennes},
  \citenamefont {Wilner}, \citenamefont {Reichman},\ and\ \citenamefont
  {Millis}}]{Kennes17}%
  \BibitemOpen
  \bibfield  {author} {\bibinfo {author} {\bibfnamefont {D.~M.}\ \bibnamefont
  {Kennes}}, \bibinfo {author} {\bibfnamefont {E.~Y.}\ \bibnamefont {Wilner}},
  \bibinfo {author} {\bibfnamefont {D.~R.}\ \bibnamefont {Reichman}}, \ and\
  \bibinfo {author} {\bibfnamefont {A.~J.}\ \bibnamefont {Millis}},\
  }\href@noop {} {\bibfield  {journal} {\bibinfo  {journal} {Nature Physics}\
  }\textbf {\bibinfo {volume} {13}},\ \bibinfo {pages} {479} (\bibinfo {year}
  {2017})}\BibitemShut {NoStop}%
\bibitem [{\citenamefont {Sentef}(2017)}]{Sentef17b}%
  \BibitemOpen
  \bibfield  {author} {\bibinfo {author} {\bibfnamefont {M.~A.}\ \bibnamefont
  {Sentef}},\ }\href {\doibase 10.1103/PhysRevB.95.205111} {\bibfield
  {journal} {\bibinfo  {journal} {Phys. Rev. B}\ }\textbf {\bibinfo {volume}
  {95}},\ \bibinfo {pages} {205111} (\bibinfo {year} {2017})}\BibitemShut
  {NoStop}%
\bibitem [{\citenamefont {Peronaci}\ \emph {et~al.}(2015)\citenamefont
  {Peronaci}, \citenamefont {Schir\'o},\ and\ \citenamefont
  {Capone}}]{Capone15}%
  \BibitemOpen
  \bibfield  {author} {\bibinfo {author} {\bibfnamefont {F.}~\bibnamefont
  {Peronaci}}, \bibinfo {author} {\bibfnamefont {M.}~\bibnamefont {Schir\'o}},
  \ and\ \bibinfo {author} {\bibfnamefont {M.}~\bibnamefont {Capone}},\ }\href
  {\doibase 10.1103/PhysRevLett.115.257001} {\bibfield  {journal} {\bibinfo
  {journal} {Phys. Rev. Lett.}\ }\textbf {\bibinfo {volume} {115}},\ \bibinfo
  {pages} {257001} (\bibinfo {year} {2015})}\BibitemShut {NoStop}%
\bibitem [{\citenamefont {Sentef}\ \emph {et~al.}(2017)\citenamefont {Sentef},
  \citenamefont {Tokuno}, \citenamefont {Georges},\ and\ \citenamefont
  {Kollath}}]{Sentef17}%
  \BibitemOpen
  \bibfield  {author} {\bibinfo {author} {\bibfnamefont {M.~A.}\ \bibnamefont
  {Sentef}}, \bibinfo {author} {\bibfnamefont {A.}~\bibnamefont {Tokuno}},
  \bibinfo {author} {\bibfnamefont {A.}~\bibnamefont {Georges}}, \ and\
  \bibinfo {author} {\bibfnamefont {C.}~\bibnamefont {Kollath}},\ }\href
  {\doibase 10.1103/PhysRevLett.118.087002} {\bibfield  {journal} {\bibinfo
  {journal} {Phys. Rev. Lett.}\ }\textbf {\bibinfo {volume} {118}},\ \bibinfo
  {pages} {087002} (\bibinfo {year} {2017})}\BibitemShut {NoStop}%
\bibitem [{\citenamefont {Oka}\ and\ \citenamefont {Aoki}(2009)}]{Oka09}%
  \BibitemOpen
  \bibfield  {author} {\bibinfo {author} {\bibfnamefont {T.}~\bibnamefont
  {Oka}}\ and\ \bibinfo {author} {\bibfnamefont {H.}~\bibnamefont {Aoki}},\
  }\href {\doibase 10.1103/PhysRevB.79.081406} {\bibfield  {journal} {\bibinfo
  {journal} {Phys. Rev. B}\ }\textbf {\bibinfo {volume} {79}},\ \bibinfo
  {pages} {081406} (\bibinfo {year} {2009})}\BibitemShut {NoStop}%
\bibitem [{\citenamefont {Kitagawa}\ \emph {et~al.}(2011)\citenamefont
  {Kitagawa}, \citenamefont {Oka}, \citenamefont {Brataas}, \citenamefont
  {Fu},\ and\ \citenamefont {Demler}}]{Kitagawa11}%
  \BibitemOpen
  \bibfield  {author} {\bibinfo {author} {\bibfnamefont {T.}~\bibnamefont
  {Kitagawa}}, \bibinfo {author} {\bibfnamefont {T.}~\bibnamefont {Oka}},
  \bibinfo {author} {\bibfnamefont {A.}~\bibnamefont {Brataas}}, \bibinfo
  {author} {\bibfnamefont {L.}~\bibnamefont {Fu}}, \ and\ \bibinfo {author}
  {\bibfnamefont {E.}~\bibnamefont {Demler}},\ }\href {\doibase
  10.1103/PhysRevB.84.235108} {\bibfield  {journal} {\bibinfo  {journal} {Phys.
  Rev. B}\ }\textbf {\bibinfo {volume} {84}},\ \bibinfo {pages} {235108}
  (\bibinfo {year} {2011})}\BibitemShut {NoStop}%
\bibitem [{\citenamefont {Haldane}(1988)}]{Haldane88}%
  \BibitemOpen
  \bibfield  {author} {\bibinfo {author} {\bibfnamefont {F.~D.~M.}\
  \bibnamefont {Haldane}},\ }\href {\doibase 10.1103/PhysRevLett.61.2015}
  {\bibfield  {journal} {\bibinfo  {journal} {Phys. Rev. Lett.}\ }\textbf
  {\bibinfo {volume} {61}},\ \bibinfo {pages} {2015} (\bibinfo {year}
  {1988})}\BibitemShut {NoStop}%
\bibitem [{\citenamefont {Dehghani}\ \emph {et~al.}(2015)\citenamefont
  {Dehghani}, \citenamefont {Oka},\ and\ \citenamefont {Mitra}}]{Dehghani15a}%
  \BibitemOpen
  \bibfield  {author} {\bibinfo {author} {\bibfnamefont {H.}~\bibnamefont
  {Dehghani}}, \bibinfo {author} {\bibfnamefont {T.}~\bibnamefont {Oka}}, \
  and\ \bibinfo {author} {\bibfnamefont {A.}~\bibnamefont {Mitra}},\ }\href
  {\doibase 10.1103/PhysRevB.91.155422} {\bibfield  {journal} {\bibinfo
  {journal} {Phys. Rev. B}\ }\textbf {\bibinfo {volume} {91}},\ \bibinfo
  {pages} {155422} (\bibinfo {year} {2015})}\BibitemShut {NoStop}%
\bibitem [{\citenamefont {Dehghani}\ and\ \citenamefont
  {Mitra}(2015)}]{Dehghani15b}%
  \BibitemOpen
  \bibfield  {author} {\bibinfo {author} {\bibfnamefont {H.}~\bibnamefont
  {Dehghani}}\ and\ \bibinfo {author} {\bibfnamefont {A.}~\bibnamefont
  {Mitra}},\ }\href {\doibase 10.1103/PhysRevB.92.165111} {\bibfield  {journal}
  {\bibinfo  {journal} {Phys. Rev. B}\ }\textbf {\bibinfo {volume} {92}},\
  \bibinfo {pages} {165111} (\bibinfo {year} {2015})}\BibitemShut {NoStop}%
\bibitem [{\citenamefont {Dehghani}\ and\ \citenamefont
  {Mitra}(2016)}]{Dehghani16}%
  \BibitemOpen
  \bibfield  {author} {\bibinfo {author} {\bibfnamefont {H.}~\bibnamefont
  {Dehghani}}\ and\ \bibinfo {author} {\bibfnamefont {A.}~\bibnamefont
  {Mitra}},\ }\href {\doibase 10.1103/PhysRevB.93.205437} {\bibfield  {journal}
  {\bibinfo  {journal} {Phys. Rev. B}\ }\textbf {\bibinfo {volume} {93}},\
  \bibinfo {pages} {205437} (\bibinfo {year} {2016})}\BibitemShut {NoStop}%
\bibitem [{\citenamefont {Uehlinger}\ \emph {et~al.}(2013)\citenamefont
  {Uehlinger}, \citenamefont {Jotzu}, \citenamefont {Messer}, \citenamefont
  {Greif}, \citenamefont {Hofstetter}, \citenamefont {Bissbort},\ and\
  \citenamefont {Esslinger}}]{Esslinger13}%
  \BibitemOpen
  \bibfield  {author} {\bibinfo {author} {\bibfnamefont {T.}~\bibnamefont
  {Uehlinger}}, \bibinfo {author} {\bibfnamefont {G.}~\bibnamefont {Jotzu}},
  \bibinfo {author} {\bibfnamefont {M.}~\bibnamefont {Messer}}, \bibinfo
  {author} {\bibfnamefont {D.}~\bibnamefont {Greif}}, \bibinfo {author}
  {\bibfnamefont {W.}~\bibnamefont {Hofstetter}}, \bibinfo {author}
  {\bibfnamefont {U.}~\bibnamefont {Bissbort}}, \ and\ \bibinfo {author}
  {\bibfnamefont {T.}~\bibnamefont {Esslinger}},\ }\href {\doibase
  10.1103/PhysRevLett.111.185307} {\bibfield  {journal} {\bibinfo  {journal}
  {Phys. Rev. Lett.}\ }\textbf {\bibinfo {volume} {111}},\ \bibinfo {pages}
  {185307} (\bibinfo {year} {2013})}\BibitemShut {NoStop}%
\bibitem [{\citenamefont {Jotzu}\ \emph {et~al.}(2014)\citenamefont {Jotzu},
  \citenamefont {Messer}, \citenamefont {Desbuquois}, \citenamefont {Lebrat},
  \citenamefont {Uehlinger}, \citenamefont {Greif},\ and\ \citenamefont
  {Esslinger}}]{Esslinger14}%
  \BibitemOpen
  \bibfield  {author} {\bibinfo {author} {\bibfnamefont {G.}~\bibnamefont
  {Jotzu}}, \bibinfo {author} {\bibfnamefont {M.}~\bibnamefont {Messer}},
  \bibinfo {author} {\bibfnamefont {R.}~\bibnamefont {Desbuquois}}, \bibinfo
  {author} {\bibfnamefont {M.}~\bibnamefont {Lebrat}}, \bibinfo {author}
  {\bibfnamefont {T.}~\bibnamefont {Uehlinger}}, \bibinfo {author}
  {\bibfnamefont {D.}~\bibnamefont {Greif}}, \ and\ \bibinfo {author}
  {\bibfnamefont {T.}~\bibnamefont {Esslinger}},\ }\href@noop {} {\bibfield
  {journal} {\bibinfo  {journal} {Nature}\ }\textbf {\bibinfo {volume} {515}},\
  \bibinfo {pages} {237} (\bibinfo {year} {2014})}\BibitemShut {NoStop}%
\bibitem [{\citenamefont {Pomarico}\ \emph {et~al.}(2017)\citenamefont
  {Pomarico}, \citenamefont {Mitrano}, \citenamefont {Bromberger},
  \citenamefont {Sentef}, \citenamefont {Al-Temimy}, \citenamefont {Coletti},
  \citenamefont {St\"ohr}, \citenamefont {Link}, \citenamefont {Starke},
  \citenamefont {Cacho}, \citenamefont {Chapman}, \citenamefont {Springate},
  \citenamefont {Cavalleri},\ and\ \citenamefont {Gierz}}]{Mitrano17}%
  \BibitemOpen
  \bibfield  {author} {\bibinfo {author} {\bibfnamefont {E.}~\bibnamefont
  {Pomarico}}, \bibinfo {author} {\bibfnamefont {M.}~\bibnamefont {Mitrano}},
  \bibinfo {author} {\bibfnamefont {H.}~\bibnamefont {Bromberger}}, \bibinfo
  {author} {\bibfnamefont {M.~A.}\ \bibnamefont {Sentef}}, \bibinfo {author}
  {\bibfnamefont {A.}~\bibnamefont {Al-Temimy}}, \bibinfo {author}
  {\bibfnamefont {C.}~\bibnamefont {Coletti}}, \bibinfo {author} {\bibfnamefont
  {A.}~\bibnamefont {St\"ohr}}, \bibinfo {author} {\bibfnamefont
  {S.}~\bibnamefont {Link}}, \bibinfo {author} {\bibfnamefont {U.}~\bibnamefont
  {Starke}}, \bibinfo {author} {\bibfnamefont {C.}~\bibnamefont {Cacho}},
  \bibinfo {author} {\bibfnamefont {R.}~\bibnamefont {Chapman}}, \bibinfo
  {author} {\bibfnamefont {E.}~\bibnamefont {Springate}}, \bibinfo {author}
  {\bibfnamefont {A.}~\bibnamefont {Cavalleri}}, \ and\ \bibinfo {author}
  {\bibfnamefont {I.}~\bibnamefont {Gierz}},\ }\href {\doibase
  10.1103/PhysRevB.95.024304} {\bibfield  {journal} {\bibinfo  {journal} {Phys.
  Rev. B}\ }\textbf {\bibinfo {volume} {95}},\ \bibinfo {pages} {024304}
  (\bibinfo {year} {2017})}\BibitemShut {NoStop}%
\bibitem [{\citenamefont {Castro~Neto}\ \emph {et~al.}(2009)\citenamefont
  {Castro~Neto}, \citenamefont {Guinea}, \citenamefont {Peres}, \citenamefont
  {Novoselov},\ and\ \citenamefont {Geim}}]{NetoRMP}%
  \BibitemOpen
  \bibfield  {author} {\bibinfo {author} {\bibfnamefont {A.~H.}\ \bibnamefont
  {Castro~Neto}}, \bibinfo {author} {\bibfnamefont {F.}~\bibnamefont {Guinea}},
  \bibinfo {author} {\bibfnamefont {N.~M.~R.}\ \bibnamefont {Peres}}, \bibinfo
  {author} {\bibfnamefont {K.~S.}\ \bibnamefont {Novoselov}}, \ and\ \bibinfo
  {author} {\bibfnamefont {A.~K.}\ \bibnamefont {Geim}},\ }\href {\doibase
  10.1103/RevModPhys.81.109} {\bibfield  {journal} {\bibinfo  {journal} {Rev.
  Mod. Phys.}\ }\textbf {\bibinfo {volume} {81}},\ \bibinfo {pages} {109}
  (\bibinfo {year} {2009})}\BibitemShut {NoStop}%
\bibitem [{\citenamefont {Roy}\ and\ \citenamefont {Juri\ifmmode \check{c}\else
  \v{c}\fi{}i\ifmmode~\acute{c}\else \'{c}\fi{}}(2014)}]{Roy14}%
  \BibitemOpen
  \bibfield  {author} {\bibinfo {author} {\bibfnamefont {B.}~\bibnamefont
  {Roy}}\ and\ \bibinfo {author} {\bibfnamefont {V.}~\bibnamefont {Juri\ifmmode
  \check{c}\else \v{c}\fi{}i\ifmmode~\acute{c}\else \'{c}\fi{}}},\ }\href
  {\doibase 10.1103/PhysRevB.90.041413} {\bibfield  {journal} {\bibinfo
  {journal} {Phys. Rev. B}\ }\textbf {\bibinfo {volume} {90}},\ \bibinfo
  {pages} {041413} (\bibinfo {year} {2014})}\BibitemShut {NoStop}%
\bibitem [{\citenamefont {Nandkishore}\ \emph {et~al.}(2012)\citenamefont
  {Nandkishore}, \citenamefont {Levitov},\ and\ \citenamefont
  {Chubukov}}]{Nandkishore12}%
  \BibitemOpen
  \bibfield  {author} {\bibinfo {author} {\bibfnamefont {R.}~\bibnamefont
  {Nandkishore}}, \bibinfo {author} {\bibfnamefont {L.}~\bibnamefont
  {Levitov}}, \ and\ \bibinfo {author} {\bibfnamefont {A.}~\bibnamefont
  {Chubukov}},\ }\href@noop {} {\bibfield  {journal} {\bibinfo  {journal}
  {Nature Physics}\ }\textbf {\bibinfo {volume} {8}},\ \bibinfo {pages} {158}
  (\bibinfo {year} {2012})}\BibitemShut {NoStop}%
\bibitem [{\citenamefont {Black-Schaffer}\ and\ \citenamefont
  {Honerkamp}(2014)}]{Black14}%
  \BibitemOpen
  \bibfield  {author} {\bibinfo {author} {\bibfnamefont {A.}~\bibnamefont
  {Black-Schaffer}}\ and\ \bibinfo {author} {\bibfnamefont {C.}~\bibnamefont
  {Honerkamp}},\ }\href@noop {} {\bibfield  {journal} {\bibinfo  {journal}
  {Journal of Physics Condensed Matter}\ }\textbf {\bibinfo {volume} {26}},\
  \bibinfo {pages} {423201} (\bibinfo {year} {2014})}\BibitemShut {NoStop}%
\bibitem [{\citenamefont {Baskaran}(2002)}]{Bhaskaran02}%
  \BibitemOpen
  \bibfield  {author} {\bibinfo {author} {\bibfnamefont {G.}~\bibnamefont
  {Baskaran}},\ }\href {\doibase 10.1103/PhysRevB.65.212505} {\bibfield
  {journal} {\bibinfo  {journal} {Phys. Rev. B}\ }\textbf {\bibinfo {volume}
  {65}},\ \bibinfo {pages} {212505} (\bibinfo {year} {2002})}\BibitemShut
  {NoStop}%
\bibitem [{\citenamefont {Black-Schaffer}\ and\ \citenamefont
  {Doniach}(2007)}]{Schaffer07}%
  \BibitemOpen
  \bibfield  {author} {\bibinfo {author} {\bibfnamefont {A.~M.}\ \bibnamefont
  {Black-Schaffer}}\ and\ \bibinfo {author} {\bibfnamefont {S.}~\bibnamefont
  {Doniach}},\ }\href {\doibase 10.1103/PhysRevB.75.134512} {\bibfield
  {journal} {\bibinfo  {journal} {Phys. Rev. B}\ }\textbf {\bibinfo {volume}
  {75}},\ \bibinfo {pages} {134512} (\bibinfo {year} {2007})}\BibitemShut
  {NoStop}%
\bibitem [{\citenamefont {D'Alessio}\ and\ \citenamefont
  {Rigol}(2014)}]{Rigol14a}%
  \BibitemOpen
  \bibfield  {author} {\bibinfo {author} {\bibfnamefont {L.}~\bibnamefont
  {D'Alessio}}\ and\ \bibinfo {author} {\bibfnamefont {M.}~\bibnamefont
  {Rigol}},\ }\href {\doibase 10.1103/PhysRevX.4.041048} {\bibfield  {journal}
  {\bibinfo  {journal} {Phys. Rev. X}\ }\textbf {\bibinfo {volume} {4}},\
  \bibinfo {pages} {041048} (\bibinfo {year} {2014})}\BibitemShut {NoStop}%
\bibitem [{\citenamefont {D'Alessio}\ and\ \citenamefont
  {Polkovnikov}(2013)}]{Alessio13}%
  \BibitemOpen
  \bibfield  {author} {\bibinfo {author} {\bibfnamefont {L.}~\bibnamefont
  {D'Alessio}}\ and\ \bibinfo {author} {\bibfnamefont {A.}~\bibnamefont
  {Polkovnikov}},\ }\href@noop {} {\bibfield  {journal} {\bibinfo  {journal}
  {Annals of Physics}\ }\textbf {\bibinfo {volume} {333}},\ \bibinfo {pages}
  {19} (\bibinfo {year} {2013})}\BibitemShut {NoStop}%
\bibitem [{\citenamefont {Abanin}\ \emph {et~al.}(2015)\citenamefont {Abanin},
  \citenamefont {De~Roeck},\ and\ \citenamefont {Huveneers}}]{Abanin15}%
  \BibitemOpen
  \bibfield  {author} {\bibinfo {author} {\bibfnamefont {D.~A.}\ \bibnamefont
  {Abanin}}, \bibinfo {author} {\bibfnamefont {W.}~\bibnamefont {De~Roeck}}, \
  and\ \bibinfo {author} {\bibfnamefont {F.~m.~c.}\ \bibnamefont {Huveneers}},\
  }\href {\doibase 10.1103/PhysRevLett.115.256803} {\bibfield  {journal}
  {\bibinfo  {journal} {Phys. Rev. Lett.}\ }\textbf {\bibinfo {volume} {115}},\
  \bibinfo {pages} {256803} (\bibinfo {year} {2015})}\BibitemShut {NoStop}%
\bibitem [{\citenamefont {Mori}\ \emph {et~al.}(2016)\citenamefont {Mori},
  \citenamefont {Kuwahara},\ and\ \citenamefont {Saito}}]{Mori16}%
  \BibitemOpen
  \bibfield  {author} {\bibinfo {author} {\bibfnamefont {T.}~\bibnamefont
  {Mori}}, \bibinfo {author} {\bibfnamefont {T.}~\bibnamefont {Kuwahara}}, \
  and\ \bibinfo {author} {\bibfnamefont {K.}~\bibnamefont {Saito}},\ }\href
  {\doibase 10.1103/PhysRevLett.116.120401} {\bibfield  {journal} {\bibinfo
  {journal} {Phys. Rev. Lett.}\ }\textbf {\bibinfo {volume} {116}},\ \bibinfo
  {pages} {120401} (\bibinfo {year} {2016})}\BibitemShut {NoStop}%
\bibitem [{\citenamefont {Bukov}\ \emph {et~al.}(2016)\citenamefont {Bukov},
  \citenamefont {Heyl}, \citenamefont {Huse},\ and\ \citenamefont
  {Polkovnikov}}]{Bukov16}%
  \BibitemOpen
  \bibfield  {author} {\bibinfo {author} {\bibfnamefont {M.}~\bibnamefont
  {Bukov}}, \bibinfo {author} {\bibfnamefont {M.}~\bibnamefont {Heyl}},
  \bibinfo {author} {\bibfnamefont {D.~A.}\ \bibnamefont {Huse}}, \ and\
  \bibinfo {author} {\bibfnamefont {A.}~\bibnamefont {Polkovnikov}},\ }\href
  {\doibase 10.1103/PhysRevB.93.155132} {\bibfield  {journal} {\bibinfo
  {journal} {Phys. Rev. B}\ }\textbf {\bibinfo {volume} {93}},\ \bibinfo
  {pages} {155132} (\bibinfo {year} {2016})}\BibitemShut {NoStop}%
\bibitem [{\citenamefont {Kuwahara}\ \emph {et~al.}(2016)\citenamefont
  {Kuwahara}, \citenamefont {Mori},\ and\ \citenamefont
  {Saito}}]{Kuwahara2016}%
  \BibitemOpen
  \bibfield  {author} {\bibinfo {author} {\bibfnamefont {T.}~\bibnamefont
  {Kuwahara}}, \bibinfo {author} {\bibfnamefont {T.}~\bibnamefont {Mori}}, \
  and\ \bibinfo {author} {\bibfnamefont {K.}~\bibnamefont {Saito}},\ }\href
  {\doibase https://doi.org/10.1016/j.aop.2016.01.012} {\bibfield  {journal}
  {\bibinfo  {journal} {Annals of Physics}\ }\textbf {\bibinfo {volume}
  {367}},\ \bibinfo {pages} {96 } (\bibinfo {year} {2016})}\BibitemShut
  {NoStop}%
\bibitem [{\citenamefont {Abanin}\ \emph {et~al.}(2017)\citenamefont {Abanin},
  \citenamefont {De~Roeck}, \citenamefont {Ho},\ and\ \citenamefont
  {Huveneers}}]{Abanin17}%
  \BibitemOpen
  \bibfield  {author} {\bibinfo {author} {\bibfnamefont {D.~A.}\ \bibnamefont
  {Abanin}}, \bibinfo {author} {\bibfnamefont {W.}~\bibnamefont {De~Roeck}},
  \bibinfo {author} {\bibfnamefont {W.~W.}\ \bibnamefont {Ho}}, \ and\ \bibinfo
  {author} {\bibfnamefont {F.~m.~c.}\ \bibnamefont {Huveneers}},\ }\href
  {\doibase 10.1103/PhysRevB.95.014112} {\bibfield  {journal} {\bibinfo
  {journal} {Phys. Rev. B}\ }\textbf {\bibinfo {volume} {95}},\ \bibinfo
  {pages} {014112} (\bibinfo {year} {2017})}\BibitemShut {NoStop}%
\bibitem [{\citenamefont {Eckardt}\ and\ \citenamefont
  {Anisimovas}(2015)}]{Eckardt15}%
  \BibitemOpen
  \bibfield  {author} {\bibinfo {author} {\bibfnamefont {A.}~\bibnamefont
  {Eckardt}}\ and\ \bibinfo {author} {\bibfnamefont {E.}~\bibnamefont
  {Anisimovas}},\ }\href {http://stacks.iop.org/1367-2630/17/i=9/a=093039}
  {\bibfield  {journal} {\bibinfo  {journal} {New Journal of Physics}\ }\textbf
  {\bibinfo {volume} {17}},\ \bibinfo {pages} {093039} (\bibinfo {year}
  {2015})}\BibitemShut {NoStop}%
\end{thebibliography}

%

\end{document}